\documentclass[aps,prd,longbibliography,secnumarabic,preprint,showpacs]{revtex4-1}
\usepackage{mathrsfs,amsmath,amssymb,bm,braket,here,comment,framed,ulem,cancel}
\usepackage{etoolbox}
\usepackage{breqn}

\usepackage{color, graphicx}
\usepackage[colorlinks=true]{hyperref}

\makeatletter
\let\cat@comma@active\@empty
\makeatother

\begin{document}


\preprint{KOBE-COSMO-18-02}
\title{Conversion of gravitons into dark photons \\
in cosmological dark magnetic fields}%

\author{Emi Masaki}%
\email[]{emi.masaki@stu.kobe-u.ac.jp}

\author{Jiro Soda}%
\email[]{jiro@phys.sci.kobe-u.ac.jp}

\affiliation{Department of Physics, Kobe University, Kobe 657-8501, Japan}%
\date{\today}



\begin{abstract}
	It is well known that gravitons can convert into photons, and {\it vice versa},
	in the presence of cosmological magnetic fields.
	We study this conversion process  in the context of the atomic dark matter scenario.
	In this scenario, we can expect cosmological dark magnetic fields, 
	which are free from the stringent constraint from
	the cosmic microwave observations.
	We find that gravitons can effectively convert into  dark photons
	in the presence of cosmological dark magnetic fields.
	The graviton--dark photon conversion
	effect may open up a new window for ultrahigh frequency gravitational waves.
\end{abstract}

\maketitle
\tableofcontents

\section{\label{intro}Introduction}

The cosmological magnetic fields provide intriguing phenomena in cosmology. 
In fact, it is known that there occurs the conversion between photons and
axions in the presence of cosmological magnetic fields,
 which can be used to probe the configuration of magnetic fields~\cite{Masaki:2017aea}. 
Remarkably, it has also been known that 
the presence of background magnetic fields induces conversion between gravitons and photons \cite{g-p1, g-p2}.
This possibility is worth investigating further.
Indeed, the graviton--photon conversion may give rise to a new perspective on the gravitational wave physics.

Apparently, the stronger the magnetic field is, the more efficient the graviton--photon conversion is.
However, the cosmological magnetic fields are constrained by observations of
the cosmic microwave background (CMB) radiations.
If we go back to the earlier universe, the strength of cosmological magnetic fields increases as the inverse of 
 the square of the scale factor.
In fact, there are some studies investigating conversion in the primordial cosmological magnetic fields 
around the recombination and in the subsequent epoch \cite{g-p3,g-p4,g-p5,g-p6,g-p7,g-p8,g-p9,g-p10}.
 The authors in \cite{g-p3,g-p4,g-p5,g-p6,g-p7,g-p8} explored the possibility that 
 CMB photons convert into gravitons.
 They proposed to utilize the deviation from the black body radiation spectrum
 as an alternative and independent probe of the cosmological magnetic fields.
In \cite{g-p9,g-p10}, they try to detect high frequency gravitational waves from primordial black holes (PBHs) \cite{source1} 
by converting it into an x ray in the cosmological background magnetic fields.
In any case, however, the conversion probability is  considerably low.
Hence, it is tempting to say that the graviton--photon conversion is irrelevant to cosmology.
However, it is still premature to conclude so.

It is well recognized that the dark matter is  one of the big puzzles of modern cosmology. 
The apparent absence of evidences for weakly interacting massive particles (WIMPs) at the LHC and from direct observations tells us that
the dark sector may contain more fertile structures to be explored~\cite{community_report,cdm1,cdm2,cdm3,cdm4}.
Among them, we focus on a dark sector charged under a hidden ${\rm U(1)}_D$ gauge symmetry in this paper.
In order to form the large scale structure of the Universe through gravitational instability, 
the dark matter has to be neutral by making atomic bound states, dubbed the atomic dark matter~\cite{adm1,adm2,adm3,adm4,adm5,adm6,adm7,Vagnozzi1,Vagnozzi2,Vagnozzi3}.
This model has been studied from various perspectives.

It should be emphasized that there exist dark photons in the atomic dark matter scenario.
Therefore,  it is natural to study graviton--dark photon conversion in this specific dark matter scenario. 
 Indeed, from the point of view of gravitational wave physics, it is worth investigating the possibility
that  gravitational waves disappear into the dark sector on the way from the source to us. 
Our main observation is that, 
from the point of ${\rm U(1)}_D$ charged dark matter, dark cosmological magnetic fields can be generated
during inflation as the conventional cosmological magnetic fields.  
In this paper, we study graviton--dark photon conversion in the presence of cosmological 
 dark magnetic fields.
Remarkably, as we will see in Section \ref{numerical}, the graviton--dark photon conversion becomes efficient.
 The main reason is that the constraint from CMB
 on the dark magnetic fields is less stringent than that 
 on the conventional cosmological magnetic fields.    
Another possible reason is that we can reduce the plasma effect by choosing the parameters in the
dark matter. Note that the graviton--dark photon conversion is not useful for
probing cosmological magnetic fields but opens up a new way to explore ultrahigh
 frequency gravitational waves. This could be possible through the photon and the dark photon mixing.

The paper is organized as follows.
In Sec. \ref{atomic_dark_matter}, we introduce the atomic dark matter scenario.
Then we review the graviton--dark photon conversion in terms of 
Schr\"odinger type formalism in Sec. \ref{g-dp}.
We calculate the conversion rate numerically in Sec. \ref{numerical}.
We also discuss implications of our results.
The final Sec. \ref{conclusion} is devoted to the conclusion.

\section{Atomic dark matter}\label{atomic_dark_matter}
 The dark sector of the Universe has not been unveiled until now.
 Hence,  there are many possible models for the dark sector~\cite{community_report,cdm1,cdm2,cdm3,cdm4}.
In this paper, we focus on the atomic dark matter model~\cite{adm1,adm2,adm3,adm4,adm5,adm6,adm7,Vagnozzi1,Vagnozzi2,Vagnozzi3}.
We consider two fermions oppositely charged under a new ${\rm U(1)}_D$ dark gauge force.
Massive fermions eventually form hydrogenlike bound states by exchanging dark photons.
 Subsequently, the structure formation begins due to the gravitational instability.
 
 In order to study the graviton--dark photon conversion, we need to understand dark plasma and dark magnetic fields.
Therefore, first, we illustrate the thermal history of the dark sector, and then we summarize parameters
which characterize the atomic dark matter scenario.
Next, we discuss a possible magnitude of cosmological dark magnetic fields.

Note that we do not intend to give a complete solution to the dark matter problem.
In fact, the study of the dark sector is now developing.
Here, we take a simple setup to illustrate our main idea.

\subsection{Thermal history of the dark sector}\label{thermal_history}
In this subsection, we review the thermal history of the dark sector in brief~\cite{thermal_relic,DM_DR,adm6}.
After the end of inflation, visible and dark sectors are  reheated and
have different temperatures, $T$ and $\hat{T}$, due to different couplings with the inflaton.
Hereafter, we denote physical quantities of the dark sector with a hat.
They could initially be equal either due to the same coupling with inflation or due to the thermal contact between them.
However, once two sectors are decoupled, entropy will be separately conserved in each sector.
Thus, in general, $T$ is different from $\hat{T}$. We parametrize the mismatch by
\begin{equation}
	\xi \equiv \frac{\hat{T}}{T}\ .
\end{equation}

Note that the energy density in the  radiation dominant period reads
\begin{equation}\label{energy_density}
	\rho_{\text{tot}} = \frac{\pi^2}{30}\,g_{*\text{tot}}(T)\,T^4\ ,
\end{equation}
where
\begin{align}
	g_{*\text{tot}} (T) &\equiv g_{*}(T) + \hat{g}_{*}(T) \notag \\
	&\equiv \sum_{\text{Bose}} g_i \left(\frac{T_i}{T}\right)^4 + \frac{7}{8} \sum_{\text{Fermi}} g_i\left(\frac{T_i}{T}\right)^4
	+ \sum_{\text{Bose}} \hat{g}_i(\hat{T})\,\xi^4(T) + \frac{7}{8} \sum_{\text{Fermi}} \hat{g}_i(\hat{T})\,\xi^4(T)\ .
\end{align}
Using the formula of the entropy per a comoving volume 
\begin{equation}
	s_{\text{tot}}=\frac{\rho_{\text{tot}}+p}{T}\ ,
\end{equation}
we obtain the entropy density in the radiation dominant period as
\begin{equation}
	s_{\rm tot} = \frac{1}{T} \frac{4}{3}\,\rho_{\text{tot}}= \frac{2\pi^2}{45}\,g_{*s,\text{tot}}(T) \,T^3\ ,
\end{equation}
where
\begin{align}
	g_{*s,\text{tot}} (T) &\equiv g_{*s}(T) + \hat{g}_{*s}(T) \notag \\
	&\equiv \sum_{\text{Bose}} g_i \left(\frac{T_i}{T}\right)^3 + \frac{7}{8} \sum_{\text{Fermi}} g_i\left(\frac{T_i}{T}\right)^3
	+ \sum_{\text{Bose}} \hat{g}_i(\hat{T})\,\xi^3(T) + \frac{7}{8} \sum_{\text{Fermi}} \hat{g}_i(\hat{T})\,\xi^3(T)\ .
\end{align}
Since the entropy per a comoving volume $s_{\text{tot}}$ conserves, the decrease of the effective degrees of freedom
 causes temperature growth against adiabatic cooling.
In the early universe, $\xi$ changes with temperature since visible sector degrees of freedom decrease with cosmological expansion.

The values of $\hat{g}_*(T)$ and $\hat{g}_{*s}(T)$ depend on how to choose a dark parameter set,
but in the following, we consider only the situation where $\hat{g}_*(T)=\hat{g}_{*s}(T)$ is established.
In other words, we assume that the dark sector is in thermal equilibrium.
Details about the parameters which fix the thermal history of the dark sector will be described in the next subsection.

Big Bang nucleosynthesis (BBN) is very sensitive to the expansion rate of the Universe
determined by the energy density~(\ref{energy_density}) through the Friedmann equation.
In fact, the number of relativistic degrees of freedom at the time of BBN is constrained 
by the abundances of the light elements.
We have a constraint in terms of the effective number of light neutrino species.
Recently, in fact, the Planck put a more stringent constraint~\cite{nu_effective_planck},
\begin{equation}\label{nu_effective}
	\Delta N_{\rm eff} = 0.11 \pm 0.23\ .
\end{equation}
This result seems to exclude the existence of relativistic particles, such as a dark photon, in the dark sector, but that is not true.
This is because the temperature of dark radiation can be much lower than a visible one.
In such a case, the percentage of $\hat{g}_*(T_{\rm BBN})$ contributing to the expansion rate is suppressed. 

Whether a dark electron/positron contributes or does not contribute depends on dark parameters;
it varies continuously from
\begin{equation}
	\hat{g}_{*}(\hat{T}_{\rm BBN}) = 2 + \frac{7}{8} \times 2\times 2 =\frac{11}{2},
\end{equation}
to
\begin{equation}
	\hat{g}_{*}(\hat{T}_{\rm BBN}) = 2\ ,
\end{equation}
where the dominant components contributing to $\hat{g}_{*}(\hat{T}_{\rm BBN})$ are dark electron/positron and dark photon, 
provided that the dark proton is massive enough to be nonrelativistic at the time of BBN.

In principle, $\hat{g}_{*}(\hat{T})$ can be calculated, once the parameters in the  dark sector are given.
In that case, the dark radiation temperature at an arbitrary $z$ is given by
\begin{equation}\label{dark_temperature}
	\hat{T}(z)=\hat{T}_0\,(z+1)\left(\frac{\hat{g}_{*} (\hat{T}_{0})}{\hat{g}_{*} (\hat{T})}\right)^{1/3}\ ,
\end{equation}
where $\hat{g}_{*} (\hat{T}_{0})=2$.
It is known that $\hat{g}_{*}(\hat{T}_{\rm BBN})=11/2$ holds for many dark parameter regions \cite{adm6},
so the temperature in both sectors rises by $(11/4)^{1/3}$ after (dark) electron/positron annihilation.
After neutrino decoupling ($\sim 1.5\,{\rm MeV}$), the degrees of freedom in both sectors change only through
 (dark) electron/positron annihilation.
 In this case,  we can make  $\xi$ constant approximately after the BBN, 
 In the following, we omit the suffix $\xi \equiv \xi_0$.
 As a side note, it is known that $\xi_0 \sim 0.5$, if the visible and dark sectors were coupled above the electroweak scale \cite{double_disk}.

 Here we provide several comments.
As far as the background cosmological expansion is concerned, the constraint on $N_{\rm eff}$ 
would be translated into the constraint on $\xi$.
Hence, dark components leave  the back ground expansion history of the Universe unchanged.
However, it affects the evolution of density fluctuations.
Since the dark radiation has coupled to the dark baryons until the dark recombination time,
dark radiation is not entirely free streaming unlike neutrinos.
Therefore, it is not straightforward to interpret the dark radiation in terms of $N_{\rm eff}$.
Using cosmological data from the CMB, baryon acoustic oscillations, and the large--scale structure,
 we can give  constraints on the strength of its interaction and the possible fraction of interacting dark matter~\cite{adm7}.
They conclude that models with eV--scale binding energy is limited to $f_{\rm int} \sim5\%$ from CMB measurements.
Here, we defined the ratio of interacting dark matter energy density $\rho_{\rm int}$ to overall dark matter energy density $\rho_{\rm DM}$ 
as
\begin{equation}\label{proportion}
	f_{\rm int} \equiv \frac{\rho_{\rm int}}{\rho_{\rm DM}}\ ,
\end{equation}
where
\begin{equation*}
	\rho_{\rm DM}=\rho_{\rm int} + \rho_{\rm CDM}\ ,
\end{equation*}
and $\rho_{\rm CDM}$ is the energy density of the collisionless dark matter.
In this paper, we set $f_{\rm int} =1$ for the sake of simplicity.
This assumption affects only a number density of the dark hydrogen atom.
The presence of charged particles which are not neutralized as the dark hydrogen atom acts to decrease the conversion probability.
Making $f_{\rm int}$ smaller than 1 has only a positive influence on conversion,
so we do not think this assumption is problematic.
In addition, in Section \ref{numerical}, we choose not the {\rm eV} scale but 10\,{\rm keV} as a binding energy.
Notice that the interacting relativistic species are also studied in \cite{cdm2,cdm4}.

\subsection{Parameters}\label{parameters}
An atomic dark matter model is a kind of hidden--charged dark matter model \cite{DM_DR,cdm1}
and behaves as the cold dark matter in the limit of large atomic binding energy and a large dark fine structure constant.
In the early universe, all of the dark atoms are ionized and in the state of the dark plasma.
When the dark radiation temperature $\hat{T}$ falls down to the binding energy of the dark atom $\hat{E}_B$,
two massive fermions start to form a hydrogenlike bound state.
We should emphasize that the thermal history of atomic dark matter is very different from that in the standard visible sector and
strongly depends on the choice of parameter sets in the dark sector.
In this subsection, we summarize parameters which characterize the atomic dark matter scenario we consider.

We shall call massless gauge boson ``dark photon,'' lighter fermion ``dark electron'' (with the mass $\hat{m}_e$), and heavier fermion ``dark proton'' (with the mass $\hat{m}_p$) in analogy with the visible sector.
In this paper, we do not get into details of the origin of the dark sector.
We assume the number of two fermions are equal and the dark sector is  neutral under the ${\rm U(1)}_D$ gauge symmetry.
A two--body system composed of a dark electron and a dark proton can be analyzed by using the reduced mass $\hat{\mu}$,
\begin{equation}\label{dark_mu}
	\hat{\mu} = \frac{\hat{m}_e\,\hat{m}_p}{\hat{m}_e+\hat{m}_p}\ .
\end{equation}
The binding energy $\hat{E}_{\rm B}$ can be expressed by the dark fine structure constant $\hat{\alpha}$ and
the reduced mass  $\hat{\mu}$ as
\begin{equation}
	\hat{E}_{\rm B} = \frac{1}{2}\,\hat{\mu}\,\hat{\alpha}^2\ .
\end{equation}
Then, denoting the mass of the finally formed dark atom as $\hat{m}_{\rm H}$, we can express the masses of dark fermions 
by imposing the following relation:
\begin{equation}\label{atom_condition}
	\hat{m}_e+\hat{m}_p-\hat{E}_{\rm B}=\hat{m}_{\rm H}\ .
\end{equation}
Solving~(\ref{dark_mu}) for $\hat{m}_e$, we obtain
\begin{equation}\label{dark_electron_mass}
	\hat{m}_e=\frac{\hat{\mu}\,\hat{m}_p}{\hat{m}_p-\hat{\mu}}\ .
\end{equation}
Substituting it into~(\ref{atom_condition}), we have the equation for $\hat{m}_p$
\begin{equation}
	\hat{m}^2_p-(\hat{m}_{\rm H}+\hat{E}_{\rm B})\,\hat{m}_p+(\hat{m}_{\rm H}+\hat{E}_{\rm B})\,\hat{\mu} =0\ .
\end{equation}
With paying an attention to the positivity of the mass $\hat{m}_p>0$, we get
\begin{equation}\label{dark_proton_mass}	
	\hat{m}_p = \frac{\hat{m}_{\rm H} +\hat{E}_{\rm B} + \sqrt{(\hat{m}_{\rm H} + \hat{E}_{\rm B})^2 - 4(\hat{m}_{\rm H} + \hat{E}_{\rm B})\hat{\mu}}}{2}\ .
\end{equation}
Now, it is easy to obtain the dark electron mass.
We identified three parameters characterizing the dark hydrogen atom, {\it i.e.}
the dark fine structure $\hat{\alpha}$, the binding energy $\hat{E}_{\rm B}$, and the dark atom mass $\hat{m}_{\rm H}$.
Even though one can choose these three parameters freely, in order to have a real solution, 
 the following condition should be satisfied
\begin{equation}
	\hat{m}_{\rm H}+\hat{E}_{\rm B} 
		\geq  4\hat{\mu}
	 = \frac{8\hat{E}_{\rm B}}{\hat{\alpha}^2} \ ,
\end{equation}
or
\begin{equation}\label{condition}	
 \frac{\hat{m}_{\rm H}}{\hat{E}_{\rm B}}\geq\frac{8}{\hat{\alpha}^2}-1\ .
\end{equation}
Recall that $\hat{T}<\hat{E}_{\rm B}$ is the condition  for the onset of the dark recombination.
In other words, $\hat{E}_{\rm B}/\xi$ fixes the redshift of dark recombination,
where we assume $\xi$ is constant (see Section \ref{thermal_history}),
\begin{equation}
	\xi \equiv \left.\frac{\hat{T}}{T}\right|_{z=0}\ .
\end{equation}
Thus, the parameter set ($\hat{\alpha}$, $\hat{E}_{\rm B}$, $\hat{m}_{\rm H}$, $\xi$)  determines the atomic dark matter scenario.
In particular, $\hat{\alpha}$ governs the interactions between dark sector components, and $\hat{m}_{\rm H}$ fixes the number density of 
atomic dark matter.

\subsection{Dark recombination}\label{sub:Dark_recombination}
Once the dark radiation temperature $\hat{T}$ falls down to the dark binding energy $\hat{E}_B$,
the dark fermions begin to recombine.
Notice that the dark recombination process strongly depends on the choice of
parameters  ($\hat{\alpha}$, $\hat{E}_{\rm B}$, $\hat{m}_{\rm H}$, $\xi$).
We refer the readers to \cite{adm6} for details of the dark recombination process.
In this paper, we narrow down our target to the parameters for which
the recombination time  is slightly shorter or comparable to the cosmological expansion time.
In fact, the standard visible sector is also categorized into this group.
We can capture the recombination process  by solving the Boltzmann equation,
\begin{equation}\label{boltzmann}
	\frac{d}{da} \hat{X}_e = \frac{\langle\widehat{\sigma v}\rangle}{aH}\left[\left(\frac{ \hat{m}_e\,\hat{m}_p}{\hat{m}_{\rm H}}\frac{\hat{T}}{2\pi}\right)^{3/2}\,e^{-\frac{\hat{E}_{\rm B}}{k_B\hat{T}}}(1-\hat{X}_e)-\hat{X}^2_e\,\hat{n}_b\right],
\end{equation}
where $X_e$ is the ionization rate of the dark electron,
\begin{equation}
	X_e \equiv \frac{\hat{n}_e}{\hat{n}_b} \equiv \frac{\hat{n}_e}{\hat{n}_e+\hat{n}_{\rm H}} 
	= \frac{\hat{n}_p}{\hat{n}_p+\hat{n}_{\rm H}}\ ,
\end{equation}
and $\hat{n}_b$, $\hat{n}_e$, and $\hat{n}_p$ are the number density of dark matter, dark electron, and dark proton, respectively.
We assume that we can neglect the number of helium atoms, and that Universe is neutral under the dark  ${\rm U(1)}_D$ charge, $\hat{n}_e=\hat{n}_p$. We also set $f_{\rm int}=1$ [see the definition of $f_{\rm int}$~(\ref{proportion})].
Note that $\langle\widehat{\sigma v}\rangle$ is the thermally averaged recombination cross section.
If an electron is directly captured to the ground state in the dark hydrogen, 
it produces a high energy photon enough to ionize other surrounding atoms.
Thus, we neglect the direct process which brings no net change,
and instead we consider  a process that an electron is captured to an excited state.
We use $\alpha^{(2)}$ to denote the thermally averaged recombination cross section excluding the direct capture to the ground states  \cite{recombination1,recombination2},
\begin{subequations}\label{cross_section}
\begin{equation}
	\langle\widehat{\sigma v}\rangle =\alpha^{(2)}=\frac{64 \pi}{\sqrt{27 \pi}}\left(\frac{\hat{\alpha}}{\hat{\mu}}\right)^2\sqrt{\frac{\hat{E}_{\rm B}}{k_B\hat{T}}}\,\phi_2\left(\frac{\hat{E}_{\rm B}}{k_B\hat{T}}\right)\ ,
\end{equation}

It is known that the function $\phi_2(\hat{E}_{\rm B}/k_B\hat{T})$ can be approximated by
\begin{equation}
	\phi_2 \left(\frac{\hat{E}_{\rm B}}{k_B\hat{T}}\right) \simeq 0.448\,\ln\left(\frac{\hat{E}_B}{k_B\hat{T}}\right)\ .
\end{equation}
\end{subequations}
Here, we should mention the applicability of the formula~(\ref{cross_section}).
In \cite{adm6}, it is pointed out that the formula ~(\ref{cross_section}) is not always applicable for
general dark parameter sets.
However, for the parameters on which we are focusing, we can adopt~(\ref{cross_section}) because
dark photons and dark fermions are kept at thermal equilibrium at a single temperature
in the following manner.
The dark electrons receive the energy frequently through the Compton scattering with a dark photon, and
the energy of dark fermions is redistributed through the Coulomb scattering.
On the other hand, for dark atoms, there is a large parameter space for which such a Compton heating does not work well at $\hat{T} \gg \hat{E}_B$.
Thus, in general, one must consider all of the mechanisms governing the energy exchange between dark photons and dark baryons,
and as a consequence ~(\ref{cross_section}) ceases to be sufficient. In the present cases, this does not happen.

The dark radiation temperature at an arbitrary $z$ is given by~(\ref{dark_temperature}).
However, we do not need to take into account
 the change of the number of relativistic species as long as we consider the period after the dark recombination.
 Therefore, we consider only the adiabatic cooling of dark radiation temperature $\hat{T}$:
\begin{equation}\label{dark_temperature2}
	\hat{T}(z) = T_0\,\xi (z+1)\ .
\end{equation}
This is because the dark recombination always happens after the dark electron/positron annihilation.

\begin{figure}[H]
\begin{center}
	\includegraphics[width=6cm]{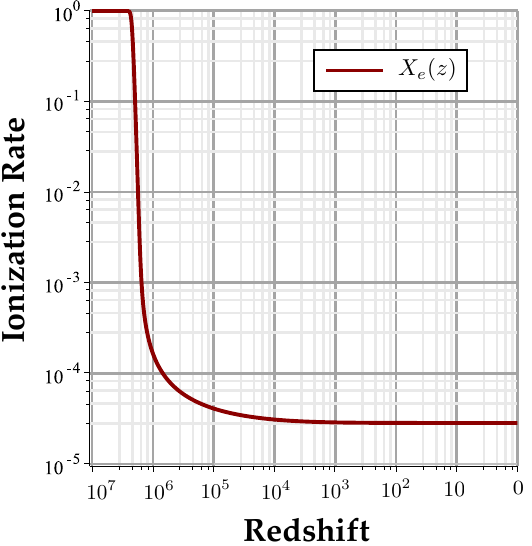}
	\caption{\label{X_e}{\bf Ionization rate.}\label{FIG_ionization_rate}\\ 
	We plot the dark recombination process as a function of redshift $z$ by solving Eq.~(\ref{boltzmann}).\\
	$\hat{\alpha} =0.05,\ \hat{E}_B =10\,{\rm keV},\ \hat{m}_{\rm H}= 1\,{\rm TeV},\ \xi = 0.5,\ \ T_0=2.73\,{\rm K}.$}
\end{center}
\end{figure}

Figure~\ref{FIG_ionization_rate} shows Eq.~(\ref{boltzmann}) solved numerically with consideration for expansion of the Universe.
Please see Section \ref{numerical} for details.

\subsection{Dark magnetic fields}\label{dark_magnetic_fields}
The Faraday rotation measurements and observations of the CMB give the upper bound for the strength of the intergalactic magnetic fields $B_{\rm IGMF} \lesssim 10^{-9}\,{\rm G}$ \cite{Magnetic_Field1,Magnetic_Field2}.
The generation mechanism has not yet been clarified, but one of the leading candidates is primordial origin.
 In the context of ${\rm U(1)}_D$ charged dark matter, dark cosmological magnetic fields can also be generated
 with the same mechanism as that for the standard cosmological magnetic fields.
Here, we consider  constraints on the dark magnetic fields.

The energy density of cosmological magnetic fields at present is
\begin{equation}
	(10^{-9}\,{\rm G})^2 \sim 8.74\times 10^{-41}\,{\rm g/cm^3}\sim1.02 \times 10^{-11}\, \rho_{cr}\,< \Omega_r\,\rho_{cr}
\end{equation}
where $\rho_{cr} =(1.91\,h^2)\times\,10^{-29}\,{\rm g/cm}^3$ is the critical density, $h=0.67$,
and $\Omega_r =9.25 \times10^{-5}$.
Since there is  no direct observation of dark magnetic fields,
a larger energy density of dark magnetic fields is allowed as long  as it does not dominant the radiation energy density.
Notice that there exist margins between radiation and cosmological magnetic field energy density.
Hence, we can take the dark magnetic field strength $\hat{B}_{0}$ at present as
\begin{equation}
	\hat{B}_{0} \sim 10^{-6}\,{\rm G}\ .
\end{equation}
In the subsequent sections, we study the effect of the presence of the dark magnetic fields.

\section{\label{g-dp}Graviton-dark photon conversion}
It is well known that gravitons can be converted into photons and {\it vice versa}, in the presence of background magnetic fields \cite{g-p1,g-p2,g-p3,g-p4,g-p5,g-p6,g-p7,g-p8,g-p9,g-p10}.
In this section, we consider graviton--dark photon conversion in the presence of background dark magnetic fields,
which was introduced in \ref{dark_magnetic_fields}.
Throughout this paper, we assume the uniform magnetic fields for the configuration of magnetic fields.
In this section, we review the conversion mechanism in terms of Schr\"odinger--like formulation.

We consider the following system:
\begin{align}\label{action}
\begin{split}
	S = &\frac{1}{16 \pi G} \int d^4x\,\sqrt{-g}\,R\\
	&+\int d^4x \sqrt{-g}\left[-\frac{1}{4}g^{\mu\rho}g^{\nu\sigma} \hat{F}_{\mu\nu}\hat{F}_{\rho\sigma}
	+\frac{\hat{\alpha}^2}{90\,\hat{m}_e^4}\left\{(\hat{F}_{\mu\nu}\hat{F}^{\mu\nu})^2+\frac{7}{4}\left(\frac{1}{2}\varepsilon_{\mu\nu\rho\sigma}\,\hat{F}^{\rho\sigma}\hat{F}^{\mu\nu}\right)^2\right\}\right],
\end{split}
\end{align}
where $g=\det(g_{\mu\nu})$, $G = 1/M^2_{\rm pl}$ is Newton's constant, $M_{\rm pl} = 1.2 \times 10^{19}\,{\rm GeV}$, and $\hat{F}_{\mu\nu} \equiv \partial_\mu\,\hat{A}_\nu-\partial_\nu\, \hat{A}_\mu$ is the field strength of the dark electromagnetic field $\hat{A}_\mu$.
We defined the dark fine structure constant $\hat{\alpha}$ and the dark electron mass $\hat{m}_e$.
The quartic terms of $\hat{F}_{\mu\nu}$ is the Euler--Heisenberg effective Lagrangian.

The action ~(\ref{action}) gives rise to the Einstein equation
\begin{align}\label{einstein}
\begin{split}
	G_{\mu\nu} =& \frac{\kappa^2}{2} \left[ g^{\alpha \beta} \hat{F}_{\alpha \mu} \hat{F}_{\beta \nu} 
	- \frac{1}{4} g_{\mu\nu} \hat{F}_{\alpha\beta} \hat{F}^{\alpha \beta}\right] \\
	&+\frac{\kappa^2}{2} \frac{\hat{\alpha}^2}{90\,\hat{m}_e^4}
	\left[ g_{\mu\nu}(\hat{F}_{\alpha\beta}\hat{F}^{\alpha\beta})^2-8\hat{F}_{\alpha\beta}\hat{F}^{\alpha\beta}g^{\rho\sigma}\hat{F}_{\rho\mu}\hat{F}_{\sigma\nu}
	-\frac{7}{4}g_{\mu\nu}\left(\frac{1}{2}\varepsilon_{\alpha\beta\rho\sigma}\,\hat{F}^{\rho\sigma}\hat{F}^{\alpha\beta}\right)^2 \right],
\end{split}
\end{align}
and the dark Maxwell equation
\begin{equation}\label{maxwell}
	\nabla_\mu\,\hat{F}^{\mu\nu}=\frac{\hat{\alpha}^2}{45\,\hat{m}_e^4} 
	\nabla_\mu\left[4(\hat{F}_{\alpha\beta}\hat{F}^{\alpha\beta}\hat{F}^{\mu\nu})+7\left(\frac{1}{2}\varepsilon_{\alpha\beta\rho\sigma}\,\hat{F}^{\rho\sigma}\hat{F}^{\alpha\beta}\frac{1}{2}\varepsilon^{\mu\nu\lambda\rho}\,\hat{F}_{\lambda\rho}\right)\right]\ .
\end{equation}

\subsection{Schr\"odinger--like formulation}
Let us start with the following metric:
\begin{equation}
	g_{\mu\nu} = \eta_{\mu\nu} + \kappa\,h_{\mu\nu}(\bm{x}, t) 
	= \eta_{\mu\alpha}[\delta^\alpha_\nu +\kappa\,h^\alpha{}_\nu(\bm{x}, t)],
\end{equation}
where $\eta_{\mu\nu}$ is the flat Minkowski metric, $\kappa \equiv  \sqrt{16 \pi G} $, and we assumed $|h_{\mu\nu}|\ll 1$.
We impose the transverse traceless (TT) gauge condition,
\begin{equation}
	h_{0\mu} = 0,\ \partial_j\,h_{ij} = 0,\ h^i{}_i = 0 \ .
\end{equation}
We can divide the dark electromagnetic field  into background and its perturbation 
\begin{align}
\begin{split}
	&\hat{F}_{\mu\nu} = \widehat{\bar{F}}_{\mu\nu} + \hat{f}_{\mu\nu} \\
	&\hat{A}_\mu = \widehat{\bar{A}}_\mu + \delta \hat{A}_\mu\ .
\end{split}
\end{align}
We use the radiation gauge,
\begin{equation}
	\hat{A}_\mu = (0, \hat{\bm{A}}_{\rm total}),\ \ \nabla \cdot \hat{\bm{A}}_{\rm total} = 0\ .
\end{equation}
The static background magnetic field and propagating electromagnetic waves are described by
\begin{align}
\begin{split}
	&\widehat{\bar{B}} \equiv \nabla \times \widehat{\bar{\bm{A}}} \\
	&\hat{F}_{0i}=\hat{E}_i=-\frac{\partial \hat{{\bm A}}_{\rm total}}{\partial t}=-\frac{\partial(\delta\hat{{\bm A}})}{\partial t}\ .
\end{split}
\end{align}
Hereafter, we denote $\widehat{\bar{B}}$ as $\hat{B}$ and $\delta \hat{\bm{A}}$ as $\hat{\bm{A}}$ for simplicity. 
We expand $h_{ij}$ and $\hat{A}_j$ by plane waves as
\begin{equation}
  \begin{cases}
    \hat{A}_j \equiv i \sum_\lambda e^{\lambda}{}_j\hat{A}_\lambda e^{- i \omega t} = i \hat{A}_+ u_j e^{-i \omega t} + i \hat{A}_\times v_j e^{- i \omega t}& \\
   h_{ij} \equiv \sum_ \lambda h_\lambda e^{\lambda}_{ij} e^{- i \omega t} = h_+ e^{+}_{ij} e^{-i \omega t} + h_\times e^{\times}_{ij} e^{-i \omega t}\ ,&
  \end{cases}
\end{equation}
where $\lambda \equiv +\ {\rm or}\ -$.
The coordinate system is set as illustrated in~Fig.~\ref{fig1}.
The linear polarization tensors for gravitational waves can be defined as
\begin{equation}
\begin{cases}
	e^+_{ij} = u_iu_j - v_iv_j \\
	e^\times_{ij} = u_iv_j + v_i u_j\ .
\end{cases}
\end{equation}
We consider monochromatic gravitational waves traveling along the $Z$ direction.
Note that we use capital $Z$ to represent the coordinate in order to avoid confusion with the redshift $z$ appearing later.
The dark magnetic field  is projected  on $x$--$y$ plane and the $y$ direction is taken along the projected magnetic field $\hat{B}$.
\begin{figure}[H]
\begin{center}
	\includegraphics[width=5cm]{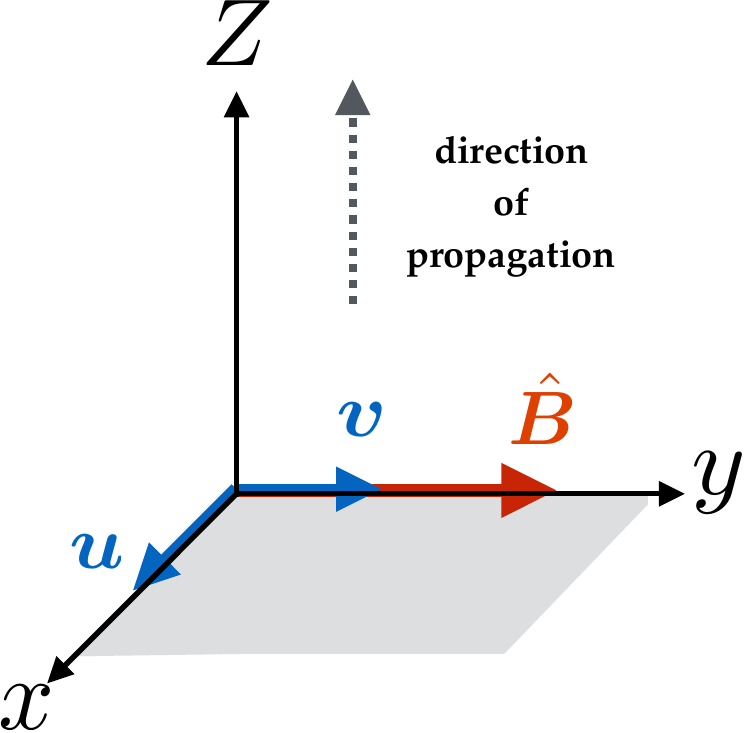}
	\caption{\label{fig2}Coordinate system}
\end{center}
\end{figure}
The linearized Einstein equation is given by
\begin{equation}\label{lin_einstein1}
	\Box h_{ij} =\kappa[\widehat{\bar{F}}_{ik}\hat{f}_{kj}+\hat{f}_{ik}\widehat{\bar{F}}_{kj}+\frac{1}{2}\eta_{ij}\hat{f}_{kl}\hat{\bar{F}}_{kl}] 
\end{equation}
and the linearized dark Maxwell equation reads
\begin{align}
	&\Box \hat{A}_i-\kappa\partial_j\left[ h_{j k} \widehat{\bar{F}}_{ki}+h_{l i} \widehat{\bar{F}}_{jl}  \right] \\
	&= 4 \varrho \hat{B}^2 \Delta \hat{A}_i +4 \varrho \hat{B}_j \hat{B}_n \partial_j(\partial_i \hat{A}_n-\partial_n \hat{A}_i)-4 \varrho \hat{B}_i \hat{B}_n \Delta \hat{A}_n+2\varrho \hat{B}^2 \Box \hat{A}_i + 7\varrho \hat{B}_m\hat{B}_i \partial_0 \partial_0 \hat{A}_m
\end{align}
where 
\begin{equation}
	\varrho\equiv \frac{4 \hat{\alpha}^2}{45 \hat{m}_e^4}\ .
\end{equation}
We obtain
\begin{equation}
\begin{cases}
	\Box h_+ = - i \kappa k \hat{B} \hat{A}_j u_j = \kappa k \hat{B} \hat{A}_+\ , \\
	\Box h_\times = - i \kappa k \hat{B} \hat{A}_j v_j = \kappa k \hat{B} \hat{A}_\times\ ,
\end{cases}
\end{equation}
by projecting~(\ref{lin_einstein1}) into $e^{+}_{ij}$
and $e^{\times}_{ij}$, respectively.
Moreover, the dark Maxwell equation is rewritten as follows:
\begin{align}\label{lin_maxwell2}
 \begin{split}
	i\sum_\lambda e^{\lambda}{}_i\,\Box \hat{A}_\lambda e^{- i \omega t} 
	= &\ i\kappa k_j h_{li} \varepsilon_{jlm} \hat{B}_m +[2 \varrho \hat{B}^2 \Box - 4 \varrho \hat{B}^2k^2]\hat{A}_i\\
	&+\left[4\,\varrho\,k^2\hat{B}-7\varrho\,\omega^2\hat{B}\right]i\hat{B}\hat{A}_\times v_ie^{-i\omega t}\ .
\end{split}
\end{align}
Projecting~(\ref{lin_maxwell2}) into $u_i$, we obtain
\begin{equation}\label{lin_maxwell3_1}
	[(1-2\varrho \hat{B}^2) \Box + 4\varrho k^2 \hat{B}^2] \hat{A}_+ = \kappa k \hat{B} h_+ \ ,
\end{equation}
and projecting~(\ref{lin_maxwell2}) into $v_i$, we obtain
\begin{equation}\label{lin_maxwell3_2}
	[(1-2\varrho \hat{B}^2)\Box + 7 \varrho k^2 \hat{B}^2]\hat{A}_\times = \kappa k \hat{B} h_\times \ .
\end{equation}
Since $\varrho^2,\ h_+ \varrho,\ h_\times \varrho$ can be neglected, ~(\ref{lin_maxwell3_1}) and ~(\ref{lin_maxwell3_2}) 
can be reduced to
\begin{equation}
 \begin{cases}
	[\Box + 4\varrho k^2 \hat{B}^2] \hat{A}_+ = \kappa k \hat{B} h_+  \\
	[\Box + 7 \varrho k^2 \hat{B}^2]\hat{A}_\times = \kappa k \hat{B} h_\times \ .
\end{cases}
\end{equation}
To sum up, we obtained  linearized Einstein and dark Maxwell equations:
\begin{equation}
 \begin{cases}
 	\Box h_+ &= \kappa k \hat{B} \hat{A}_+ \\
	[\Box + 4\varrho k^2 \hat{B}^2] \hat{A}_+ &= \kappa k \hat{B} h_+ \\
	\Box h_\times &= \kappa k \hat{B} \hat{A}_\times \\
	[\Box + 7 \varrho k^2 \hat{B}^2]\hat{A}_\times &= \kappa k \hat{B} h_\times\ .
\end{cases}
\end{equation}
Assuming $\omega \simeq k$ and $\Box = (\omega + i \partial_Z)(\omega - i \partial_Z) = (\omega + i \partial_Z)(\omega + k) \simeq 2\omega (\omega + i \partial_Z)$, we can simplify the equations as 
\begin{align*}
	(\omega + i \partial_Z) h_\lambda &= 2\sqrt{\pi} \frac{\hat{B}}{M_{\rm pl}} \hat{A}_\lambda \\
	(\omega + i \partial_Z)\hat{A}_\lambda + \frac{\beta}{2}\,\varrho\,\omega \hat{B}^2 \hat{A}_\lambda &= 2 \sqrt{\pi} \frac{\hat{B}}{M_{\rm pl}}  h_\lambda \ ,
\end{align*}
where $\beta=4\,(\text{for}\ \lambda=+)$ or $\beta=7\,(\text{for}\ \lambda=\times)$.
By introducing $\Psi$,
\begin{equation*}
  \Psi\ \equiv\ \left(
	\begin{array}{c}
		h_+(Z)\\
		\hat{A}_+(Z)\\
		h_\times(Z) \\
		\hat{A}_\times(Z)\\
	\end{array}
   \right) e^{-i\omega Z } \ ,
\end{equation*}
we can finally deduce the basic Schr\"odinger type equation 
\begin{equation}
	i \dfrac{d}{dZ}
	\Psi
	=
	\left(	\begin{array}{cccc}
		0&\hat{\Delta}_{g\gamma}&0&0 \\
		\hat{\Delta}_{g\gamma}&\hat{\Delta}_\gamma&0&0\\
		0 &0 &0& \hat{\Delta}_{g\gamma}\\
		0 &0 &\hat{\Delta}_{g\gamma}&\hat{\Delta}_\gamma \\
	\end{array}\right)
	\Psi\ ,
\end{equation}
where the mixing term is defined by
\begin{subequations}\label{Delta}
\begin{equation}\label{Delta_mix}
	\hat{\Delta}_{g\gamma} \equiv 2\sqrt{\pi} \frac{\hat{B}}{M_{\rm pl}}\ ,
\end{equation}
and the effective photon mass term is given by
\begin{equation}\label{Delta_gamma}
	\hat{\Delta}_{\gamma} \equiv \hat{\Delta}_p + \hat{\Delta}_{\rm QED}\ .
\end{equation}
The dispersion relation for dark electromagnetic waves propagating in the dark plasma is modified to 
\begin{equation*}
	\omega^2 = k^2 + \hat{\omega}^2_p\ ,
\end{equation*}
where the plasma frequency is defined by
\begin{equation}
	\hat{\omega}^2_p = 4\pi\hat{\alpha}\frac{\hat{n}_e}{\hat{m}_e}\ ,
\end{equation}
with the dark electron number density $\hat{n}_e$.
Thus, the effect of the plasma is described by
\begin{equation}\label{Delta_p}
	\hat{\Delta}_p \equiv \frac{\hat{\omega}^2_p}{2\omega}\ .
\end{equation}
The QED effect \cite{QED} depends on the polarization $\lambda$,
\begin{equation}\label{Delta_QED}
	\hat{\Delta}_{\rm QED\,+} \equiv - 2\,\varrho\,\omega \hat{B}^2 \ ,\quad 
	\hat{\Delta}_{\rm QED\,\times} \equiv - \frac{7}{2}\,\varrho\,\omega \hat{B}^2\ .
\end{equation}
\end{subequations}
These effects~(\ref{Delta_gamma}) --~(\ref{Delta_QED}) on gravitational waves need not be considered, 
because the interaction of gravitational waves with the medium is very weak.
Note that the Schr\"odinger--like equation can be block diagonalized as 
\begin{equation}
	i \frac{d}{dZ} \Psi = M_{\rm mix} \Psi \ ,
\end{equation}
where the mixing matrix $M_{\rm mix}$ is defined as follows:
\begin{equation}
	M_{\rm mix}
	\equiv
	\left(
	\begin{array}{cc}
	M & \bm{0}\\
	\bm{0} & M\\
	\end{array}
	\right),
	\ \ 
	M
	\equiv
	\left(
	\begin{array}{cc}
	0 & \hat{\Delta}_{g\gamma}\\
	\hat{\Delta}_{g\gamma} & \hat{\Delta}_{\gamma}\\
	\end{array}
	\right).
\end{equation}
We see that each of two independent polarization components of gravitational waves mixes with a particular polarization 
component of electromagnetic waves.
Mathematically, we obtained the equations with 
the similar structure to the conversion between photons and axions \cite{g-p2}.
However, there is a qualitative difference. In the case of photon to axion conversion,
when an unpolarized electromagnetic wave propagates through homogeneous magnetic fields, the linear polarization appears,
because only one component can convert into axions.
In the case of photon--graviton conversion, cases like that never happen.

We can discuss  each polarization separately.
Thus, we analyze the time evolution by reduced equation
\begin{equation}\label{schrodinger}
	i \frac{d}{dZ}\,\psi = M \psi\ ,
\end{equation}
where
\begin{equation*}
  \psi\ \equiv\ \left(
	\begin{array}{c}
		h_\lambda(Z)\\
		\hat{A}_\lambda(Z)
	\end{array}
   \right) e^{-i\omega Z } \ .
\end{equation*}
Now, we derive the conversion probability by solving~(\ref{schrodinger}).
To this end, we introduce an orthogonal matrix $O$ which diagonalizes $M$
\begin{equation}
	OMO^\dag 
	=
	 \left(
    \begin{array}{cc}
    \lambda_+ & 0 \\
   0 & \lambda_-\\
    \end{array}
    \right),\ \ 
    O
    =
     \left(
    \begin{array}{cc}
   \cos \theta & \sin \theta \\
   -\sin \theta & \cos \theta\\
    \end{array}
    \right)\ ,
\end{equation}
where 
the eigenvalues of $M$ are given by
\begin{equation}
	\lambda_\pm = \frac{\hat{\Delta}_\gamma \pm \sqrt{(\hat{\Delta}_\gamma)^2+(2\hat{\Delta}_{g\gamma})^2}}{2}\ ,
\end{equation}
and the definition of mixing angle $\theta$ is given by
\begin{align}
\begin{split}
	&\tan 2 \theta = - \frac{2 \hat{\Delta}_{g\gamma}}{\hat{\Delta}_\gamma} \ ,\quad 
	\cos2\theta 
	= -\frac{\hat{\Delta}_\gamma}{\hat{\Delta}_{osc}}\ ,\quad 
	\sin2\theta = \frac{2 \hat{\Delta}_{g\gamma}}{\hat{\Delta}_{osc}}\ ,
\end{split}
\end{align}
where we defined the oscillation length $\hat{\Delta}_{\rm osc}^{-1}$
\begin{equation}
	\hat{\Delta}_{\rm osc} \equiv \lambda_+ -\lambda_-= \sqrt{(\hat{\Delta}_\gamma)^2+(2\hat{\Delta}_{g\gamma})^2}\ .
\end{equation}
Defining $\tilde{\psi}\equiv O\,\psi$, we can solve~(\ref{schrodinger}) as
\begin{equation}
	\tilde{\psi}_i(Z)\ =\ \tilde{\psi}_i(Z_0)\,e^{-i \lambda_i Z}\ .
\end{equation}
Thus, we have
\begin{equation}
	\psi_i(Z)\ =\ \sum_{j=1}^2 O^{\dag}_{ij} \tilde{\psi}_j
	\ =\ \sum_{j=1}^2 O^{\dag}_{ij}[O \psi(Z_0)]_j e^{-i \lambda_j Z}\ ,
\end{equation}
where $\lambda_1\equiv \lambda_+$, and $\lambda_2\equiv \lambda_-$.
Finally, we obtain
\begin{align}\label{ans_1}
\begin{split}
	h_\lambda(Z) &= [\cos^2 \theta h_\lambda(Z_0)+ \cos\theta\sin\theta \hat{A}_\lambda(Z_0)]e^{-i\lambda_+ Z}
	+[\sin^2 \theta h_\lambda(Z_0)- \cos\theta\sin\theta \hat{A}_\lambda(Z_0)]e^{-i \lambda_- Z} \\
	\hat{A}_\lambda (Z)&= [\cos\theta\sin\theta h_\lambda(Z_0)+\sin^2\theta \hat{A}_\lambda(Z_0)]e^{- i \lambda_+ Z} 
	+ [- \cos\theta \sin\theta h_\lambda (Z_0)+ \cos^2 \theta \hat{A}_\lambda(Z_0)]e^{-i \lambda_- Z}\ .
\end{split}
\end{align}
Assuming the initial conditions $h_\lambda(Z_0)=1, \hat{A}_\lambda (Z_0) = 0$, we find
\begin{align*}
\begin{split}
	h_\lambda(Z) &= \cos^2 \theta e^{-i\lambda_+ Z}+\sin^2 \theta e^{-i \lambda_- Z} \ ,\\
	\hat{A}_\lambda (Z)&= \cos\theta\sin\theta e^{- i \lambda_+ Z} - \cos\theta \sin\theta e^{-i \lambda_- Z} \ .
\end{split}
\end{align*}
Thus, the conversion probability after propagating the distance $Z$ can be deduced as
\begin{equation}\label{rate}
	P = \left(\frac{2 \hat{\Delta}_{g\gamma}}{\hat{\Delta}_{osc}} \right)^2 \sin^2 \left(\frac{\hat{\Delta}_{osc}}{2} Z\right)\ .
\end{equation}
The effect of photon effective mass $\hat{\Delta}_{\gamma}$ can vanish,
since $\hat{\Delta}_p$ and $\hat{\Delta}_{\rm QED}$ have opposite signs.
In this case, the conversion probability ceases to depend on the frequency $\omega$, and the probability becomes 
\begin{equation}\label{rate2}
	P = \sin^2 (\hat{\Delta}_{g\gamma}\,Z)\ .
\end{equation}
This corresponds to the maximum mixing $\theta = \pi/4$ for which the complete conversion is possible.
However, the typical value of the mixing term is 
\begin{equation}
	\hat{\Delta}_{g \gamma} \equiv 2\sqrt{\pi} \frac{\hat{B}}{M_{\rm pl}} = 9.04 \times 10^{-7}\,{\rm Mpc}^{-1}\,\left( \frac{1.2 \times 10^{19} {\rm GeV}}{M_{\rm pl}} \right) \left(\frac{\hat{B}}{10^{-6}\,{\rm G}}\right)\ .
\end{equation}
Hence, in order to achieve  $\hat{\Delta}_{g\gamma}\,Z \sim \pi/2$, we need $Z \sim 10^{12}\,{\rm pc}$.
Therefore, the efficient conversion requires strong dark magnetic fields  even in the absence of $\hat{\Delta}_{\gamma}$.

In the case that the cosmic expansion cannot be neglected,
we need to take into account the time evolution of physical quantities and solve the equation

\begin{align}\label{eq:sch_expansion}
i\dfrac{d}{da}
    \left(\begin{array}{c}
       h_\lambda(a)\\
       A_\lambda(a)\\
     \end{array}\right)
    =
    \left(
    \begin{array}{cc}
    0 & \Delta_{g \gamma}(a)/aH \\
    \Delta_{g \gamma}(a)/aH& \Delta_\gamma(a)/aH\\
    \end{array}
    \right)
     \left(\begin{array}{c}
       h_\lambda(a)\\
       A_\lambda(a)\\
     \end{array}\right)\ ,
\end{align}
where $a$ is the scale factor, $H$ is Hubble parameter
and we used the following relation:
\begin{equation}
	\frac{d}{dZ}=\frac{d}{dt} = aH\frac{d}{da}\ .
\end{equation}
Equation~(\ref{eq:sch_expansion}) can only be solved analytically in limited circumstances.

\section{The conversion rate: Numerical results}\label{numerical}
In the previous sections, it turned out that
the gravitational waves and the dark photon can mix with each other,
if dark magnetic fields exist in the context of atomic dark matter.
In this section, we set source of gravitational waves and atomic dark matter scenario concretely.
Then, we investigate the conversion probability with numerical calculations.

We suppose gravitational waves are emitted from PBHs.
After inflation, very light primordial black holes with the mass $<10^8\,{\rm g}$ can dominate~\cite{source1}.
Indeed, high energy gravitational waves can be emitted during evaporation of PBHs before the BBN.
The mass of PBHs determines the peak frequency 
and the maximum value can be {\rm MeV} today.
The density parameter can be $h^2_0\Omega_{\rm gw} \sim 10^{-8}$.
It is expected that such  high energy gravitational waves can be detectable by converting it into an x ray 
in the presence of cosmological magnetic fields~\cite{g-p9,g-p10}.

We are interested in the conversion probability in the cosmological history.
The dark recombination occurs before the recombination in the standard visible sector.
 The dark electron number density drastically drops off around that epoch.
We derive the conversion probability by solving the equations starting at $z=10^5$ 
where the dark electron density is almost fixed (see~Fig.~\ref{X_e}).

We solve Eqs.~(\ref{eq:sch_expansion}) and ~(\ref{boltzmann}), with the parameters
\begin{equation}\label{dark_parameter}
	\xi = 0.5,\ \hat{\alpha} =0.05,\ \hat{E}_B =10\,{\rm keV},\ \hat{m}_{\rm H}= 1\,{\rm TeV}\ .
\end{equation}
Using~(\ref{dark_mu}),~(\ref{dark_electron_mass}), and ~(\ref{dark_proton_mass}), we obtain
\begin{equation}
	\hat{m}_e = 8\,{\rm MeV},\ \ \hat{m}_p=1\,{\rm TeV},\ \ \hat{\mu} = 8\,{\rm MeV}\ \ .
\end{equation}
In the following, we adopt the normalization $a_0 =1$ at redshift $z_0=0$, and represent a physical quantity of this age 
with a suffix 0 added.
The dark baryon density (dark hydrogen density) is given by
\begin{subequations}\label{i_prameters}
\begin{equation}
	\hat{n}_{b,0} \equiv \frac{\rho_{cr} \cdot \Omega_{\rm CDM}}{\hat{m}_{\rm H}}\ ,
\end{equation}
where we assumed $\Omega_{\rm CDM} =0.267$ and $f_{\rm int}=1~(\ref{proportion}).$
Note that the present CMB temperature is 
\begin{equation}
T_0 = 2.73\,{\rm K}\ .
\end{equation}
As previously mentioned in Section \ref{dark_magnetic_fields},  we set
\begin{equation}
\hat{B}_{0} = 1.95\times10^{-8}\,{\rm eV}^2,
\end{equation}
where we used 1\,G $=1.95\times10^{-2}\,\rm{eV}^2$.
The wavelength of gravitational waves at present is stretched by redshift, so its energy is given by
\begin{equation}
	\omega_0 = 1\,{\rm eV}\ .
\end{equation}
\end{subequations}

We consider the time evolution of each parameter in~(\ref{eq:sch_expansion}) and~(\ref{boltzmann}).
The Hubble parameter is given by 
\begin{equation}
	H\equiv H_0 \sqrt{\frac{\Omega_m}{a^3}+\frac{\Omega_r}{a^4}+\Omega_\Lambda}\ ,
\end{equation}
where we used the density parameters normalized by the present ($z=0$) critical density $\rho_{cr} = (8.1\cdot h^2)\times 10^{-11}\,{\rm eV}^4$.
More precisely, we set $h=0.67$, $\Omega_m = 0.315$, $\Omega_r=9.245\times10^{-5}$, $\Omega_\Lambda = 0.685$.
The Hubble parameter and scale factor always appear with the combination,
\begin{subequations}
\begin{equation}
	aH = H_0 \sqrt{\frac{\Omega_m}{a}+\frac{\Omega_r}{a^2}+a^2\Omega_\Lambda}\ .
\end{equation}
The dark electron number density is given by
\begin{equation}
	\hat{n}_e (a) = X_e (a)\,\frac{\hat{n}_{b,0}}{a^3}\ ,
\end{equation}
and the dark radiation temperature reads
\begin{equation}
	\hat{T}(a)=\frac{T_{0}\,\xi}{a}\ .
\end{equation}

Since the strength of the dark magnetic field evolves as the inverse of the scale factor squared,
 the frequency of gravitational waves evolves as the inverse of the scale factor,
and the number density of the dark electron evolves as the inverse of the  scale factor cubic,
the variables $\hat{\Delta}$ introduced in ~(\ref{Delta}) evolve as follows:
\begin{equation*}
	\hat{\Delta}_{g\gamma} \equiv 2\sqrt{\pi} \frac{\hat{B}}{M_{\rm pl}} \propto \frac{1}{a^2} \tag{\ref{Delta_mix}} \ ,
\end{equation*}
\begin{equation*}
	\hat{\Delta}_p \equiv \frac{\hat{\omega}^2_p}{2\omega} \tag{\ref{Delta_p}} 
	=\frac{4\pi\hat{\alpha}}{2\omega}\frac{\hat{n}_e}{\hat{m}_e}\propto \frac{1}{a^2} \ ,
\end{equation*}
\begin{equation*}
	\hat{\Delta}_{\rm QED} \propto - \,\varrho\,\omega \hat{B}^2 
	\propto -\frac{4 \hat{\alpha}^2}{45 \hat{m}_e^4}\,\omega \hat{B}^2\tag{\ref{Delta_QED}}\propto \frac{1}{a^5}\ .
\end{equation*}
Thus, we have the following equations:
\begin{align}
	&\frac{\hat{\Delta}_{g\gamma}(a)}{aH} =\frac{1}{H}\frac{\hat{\Delta}_{g\gamma,0}}{a^3}\\
	&\frac{\hat{\Delta}_{\gamma}(a)}{aH} =\frac{1}{H}\left[\frac{2\pi\hat{\alpha}}{\hat{m}_e}\frac{\hat{n}_{b,0}}{\omega_0}\frac{X_e(a)}{a^3}+\frac{\hat{\Delta}_{{\rm QED},0}}{a^6}\right]\ .
	\end{align}
\end{subequations}
In ~Fig.~\ref{X_e}, we show the dark recombination process.
Note that we do not consider dark reionization.

\begin{figure}[H]
\begin{center}
	\includegraphics[width=13cm]{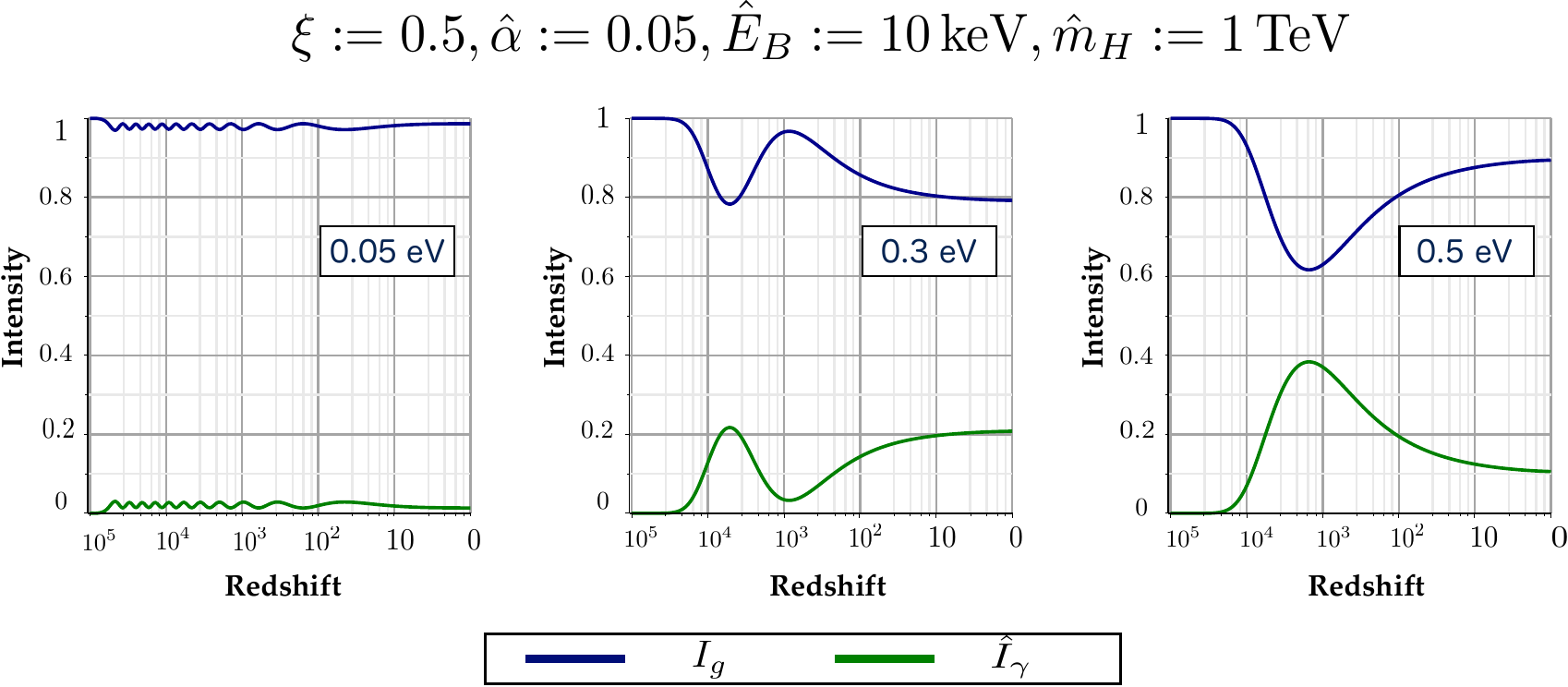}\label{FIG2}
	\caption{\label{dark_conversion1}{\bf Graviton -- dark photon conversion ($\hat{B}_{0} =1\,\mu{\rm G}$).}\\ The intensity of graviton and 
	dark photon is plotted as a function of redshift $z$, changing current gravitational waves energy $\omega_0$.}
\end{center}
\end{figure}

\begin{figure}[H]
\begin{center}
	\includegraphics[width=13cm]{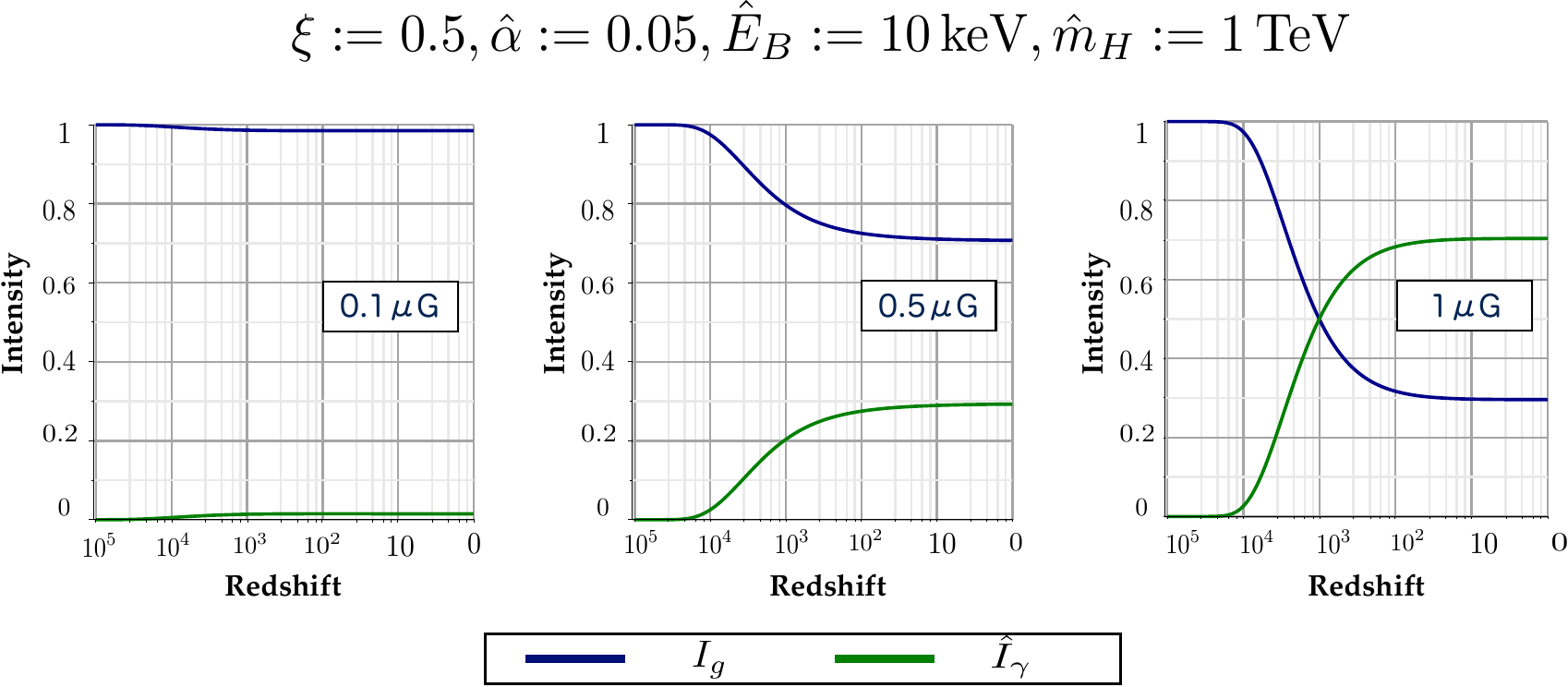}\label{FIG3}
	\caption{\label{dark_conversion2}{\bf Graviton -- dark photon conversion ($\omega_0 =1\,{\rm eV}$).}\\ The intensity of graviton and 
	dark photon is plotted as a function of redshift $z$, changing current dark magnetic field strength $\hat{B}_{0}$.}
\end{center}
\end{figure}

We solve Eq.~(\ref{eq:sch_expansion}) numerically from $z=10^5$ where dark neutralization is sufficiently advanced to $z=0$
 with the initial conditions
\begin{equation}
	I_g (z=10^5) =1,\ \hat{I}_\gamma (z=10^5) = 0\ .
\end{equation}
In ~Fig.~\ref{dark_conversion1}, we repeated the calculations by changing
the current energy of gravitational waves $\omega_0$ while fixing the current dark magnetic field strength $\hat{B}_{0}=1\,\mu{\rm G}$.
In ~Fig.~\ref{dark_conversion2}, we did the same by changing the current dark magnetic field strength  $\hat{B}_{0}$ while fixing
the current energy of gravitational waves $\omega_0=1\,{\rm eV}$.
Both ~Fig.~\ref{dark_conversion1} and ~Fig.~\ref{dark_conversion2} show that
graviton--dark photon conversion can be effective, if dark magnetic fields are larger than the conventional one
and such high frequency gravitational waves exist.


\section{\label{conclusion}Conclusion}
We studied the graviton--dark photon conversion in the presence of the cosmological dark magnetic fields
in the scenario of the atomic dark matter and found the conversion can be effective.
This is in contrast to the graviton--photon conversion in the conventional magnetic fields,
which is less efficient due to the Planck mass suppression 
and the upper bound for the cosmological magnetic field.
In the present case, since there is no robust constraint for the dark cosmological magnetic fields 
and the choice of the dark parameter set, 
 the probability of graviton--dark photon conversion can be high in the atomic dark matter scenario.
 It should be mentioned that
the graviton--dark photon conversion is useful as a detector for ultrahigh frequency gravitational waves.
This can be realized through the photon and dark photon mixing although the detailed method is model dependent.

It has been argued that gravitational waves from the PBHs can be observed
by converting them into x rays in the cosmological magnetic fields~\cite{g-p9,g-p10,source1}.
However, our results suggest the possibility that, within the atomic dark matter scenario, such an observation method should be reconsidered
by taking into account the graviton--dark photon conversion.

There are remaining problems in the study of the graviton--dark photon conversion.
In the present work, we found a dark parameter set which shows the efficient graviton--dark photon conversion.
Although the graviton--dark photon conversion occurs at the very high frequency in the cases we found, 
there is still  a chance to find a dark parameter set for which the conversion
 is effective even for lower frequency regions.
Indeed, we can choose freely $\hat{\alpha}$, $\hat{E}_{\rm B}$, and $\hat{m}_{\rm H}$ as long as they satisfy condition~(\ref{condition}).
In Appendix \ref{app:various}, we presented other numerical results Figs.~\ref{FIG5}--\ref{FIG20}.
However, formula~(\ref{cross_section}) is not always valid for these parameters. In fact,
we should scrutinize the recombination process in detail~\cite{adm6}.
If we could find a dark parameter set for the effective conversion of the low frequency gravitational waves,
we may be able to observe the dark photon conversion into gravitational waves.
Moreover,  we may be able to use gravitational waves to explore the atomic dark matter.
We leave these issues for future work.

\acknowledgments
E.M. was in part supported by JSPS KAKENHI Grant Number JP 18J20018.
J.S. was in part supported by JSPS KAKENHI Grant Numbers JP18H04589,
JP17H02894, JP17K18778, JP15H05895, JP17H06359.
We are also supported by JSPS Bilateral Joint Research
Projects (JSPS-NRF Collaboration) “String Axion Cosmology”.


\newpage

\appendix


\section{Density matrix formulation of conversion}
There is a possibility that a dark photon would be attenuated by dark Thomson scattering \cite{g-p9}.
This phenomenon can be described by the following formalism.
We confirmed numerically that the same result can be obtained with both Schr\"odinger--like and density matrix formulation.
Since the same result was obtained using two kinds of expressions, it assures the correctness of our numerical calculation but it does not affect the contents.

It is possible to follow the time evolution of the system in terms of 
an equation for the density matrix $\rho$ \cite{decoherence},
\begin{equation}\label{Liouville}
	i\frac{d\rho}{dZ} = M_{\rm tot}\rho-\rho M^\dag_{\rm tot}\ ,
\end{equation}
where
\begin{align}
	\rho(Z)\ &\equiv\ \left(\begin{array}{c}
	h_\lambda(Z)\\       
	A_\lambda(Z)\\
	\end{array}\right)
	\otimes \left(h_\lambda^*(Z)\ A_\lambda^*(Z)\right) \equiv 
	\left(
	\begin{array}{cc}
	I_g & K-iL\\
	K+iL & \hat{I}_\gamma\\
	\end{array}
	\right)\ .
\end{align}
When $M_{\rm tot}$ is Hermitian and $M_{\rm tot}=M$, the right--hand side of ~(\ref{Liouville}) is represented by a commutator
\begin{equation}\label{Liouville2}
	i\frac{d\rho}{dZ} = [M_{\rm tot},\rho]\ ,
\end{equation}
Note that Eqs.~(\ref{schrodinger}) and (\ref{Liouville2}) are completely equivalent.
Solving the above equation formally, we obtain
\begin{equation}
	\rho (Z) = e^{-iM_{\rm tot}Z}\rho(0)\,e^{iM_{\rm tot}Z}\ .
\end{equation}
However, there is no actual isolated system. Indeed, any system generically interacts with its environment.
The interaction with the environment makes $M_{\rm tot}$ non--Hermitian, and the conversion probability is not conserved.
Thus the density matrix formalism~(\ref{Liouville}) is more general.

The total Hamiltonian of an open system is given by 
\begin{equation}\label{M_tot}
	M_{\rm tot} = M - i \Gamma\ ,
\end{equation}
where $M$ and the damping factor $\Gamma$ are both Hermitian.
For example, in the case of dark Thomson scattering
\begin{equation}
	\Gamma = \left(
	\begin{array}{cc}
	0 & 0\\
	0 & \hat{\sigma}_{\rm T}\hat{n}_e\\
	\end{array}
	\right) .
\end{equation}
Substitution of~(\ref{M_tot}) into Eq.~(\ref{Liouville}) leads to
\begin{equation}\label{Liouville3}
	i\frac{d\rho}{dZ} = [M,\rho]-i\{\Gamma,\rho\}\ .
\end{equation}
The first term in~(\ref{Liouville3}) represents the usual Schr\"odinger term, and the second one describes the decoherence.

In the case that the cosmic expansion cannot be neglected, 
we  need to take into account the time evolution of physical quantities and solve the equation
\begin{equation}\label{Liouville4}
	i aH\,\frac{d \rho}{da} = [M(a), \rho]-i\{\Gamma(a), \rho\}\ .
\end{equation}
Here, we used the following relation:
\begin{equation}
	\frac{d}{dZ}=\frac{d}{dt} = aH\frac{d}{da}\ ,
\end{equation}
where  $a$ is the scale factor and $H$ is the Hubble parameter.
Each component of Eq.~(\ref{Liouville4}) can be written explicitly as follows:
\begin{equation}\label{conversion}
	\dfrac{d}{da}
	\left(\begin{array}{c}
		I_g\\
		\hat{I}_\gamma\\
		K\\
		L\\
	\end{array}\right)
	=
	\left(	\begin{array}{cccc}
		0&0&0&2\hat{\Delta}_{g\gamma}(a)/aH \\
		0&- \Gamma(a)/aH&0&-2\hat{\Delta}_{g\gamma}(a)/aH\\
		0 &0 &-\Gamma(a)/aH&\hat{\Delta}_{\gamma}(a){/aH}\\
		-\hat{\Delta}_{g\gamma}(a)/aH &\hat{\Delta}_{g\gamma}(a)/aH&-\hat{\Delta}_{\gamma}(a){/aH}&-\Gamma(a)/aH \\
	\end{array}\right)
	\left(\begin{array}{c}
		I_g\\
		\hat{I}_\gamma\\
		K\\
		L\\
	\end{array}\right)\ .
\end{equation}


\section{Graviton--dark photon conversion with various parameters}\label{app:various}
\subsection{Change in dark fine structure constant $\hat{\alpha}$}
\begin{figure}[H]
\begin{center}
	\includegraphics[width=13cm]{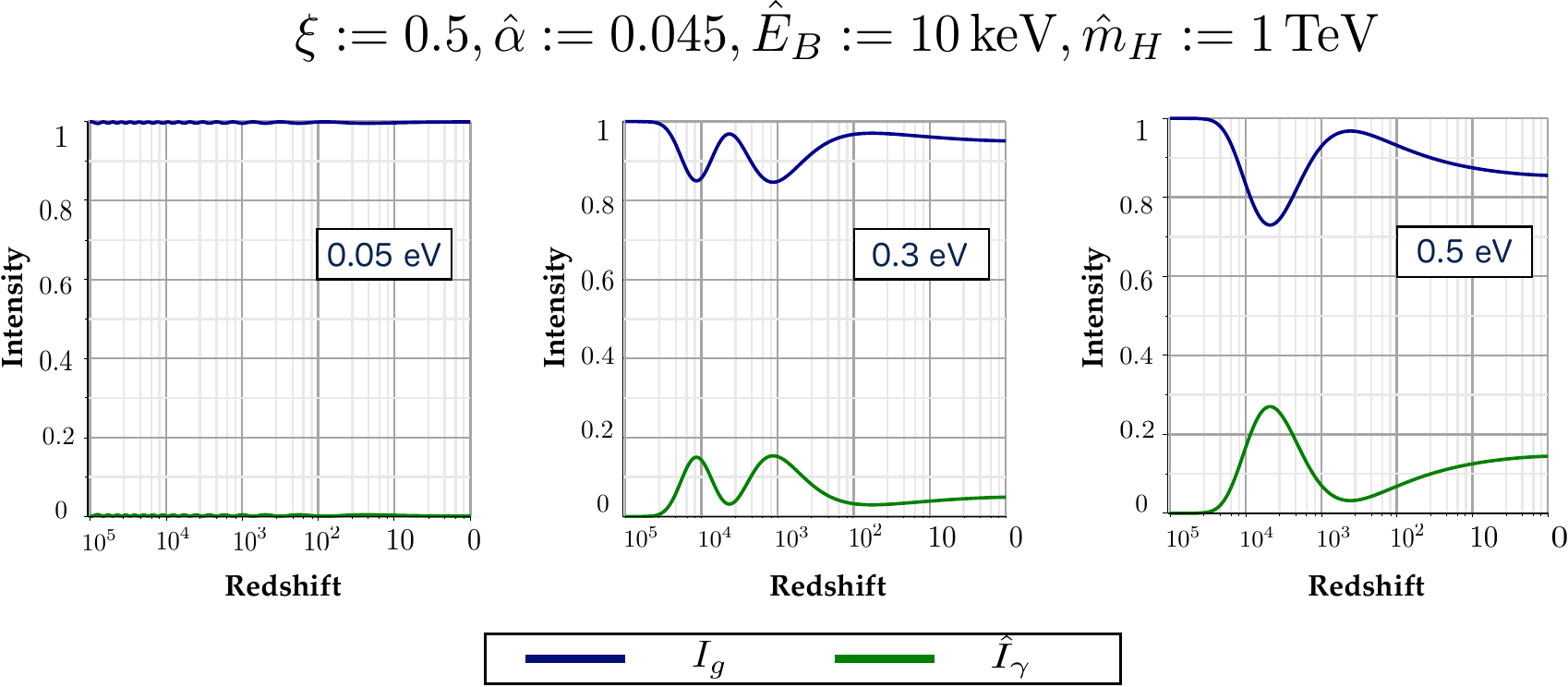}
	\caption{\label{FIG5}{\bf Smaller $\hat{\alpha}\ $($\hat{B}_{0} =1\,\mu{\rm G}$)}\\ The intensity of graviton and 
	dark photon is plotted as a function of redshift $z$, changing current gravitational waves energy $\omega_0$.}
\end{center}
\end{figure}
\begin{figure}[H]
\begin{center}
	\includegraphics[width=13cm]{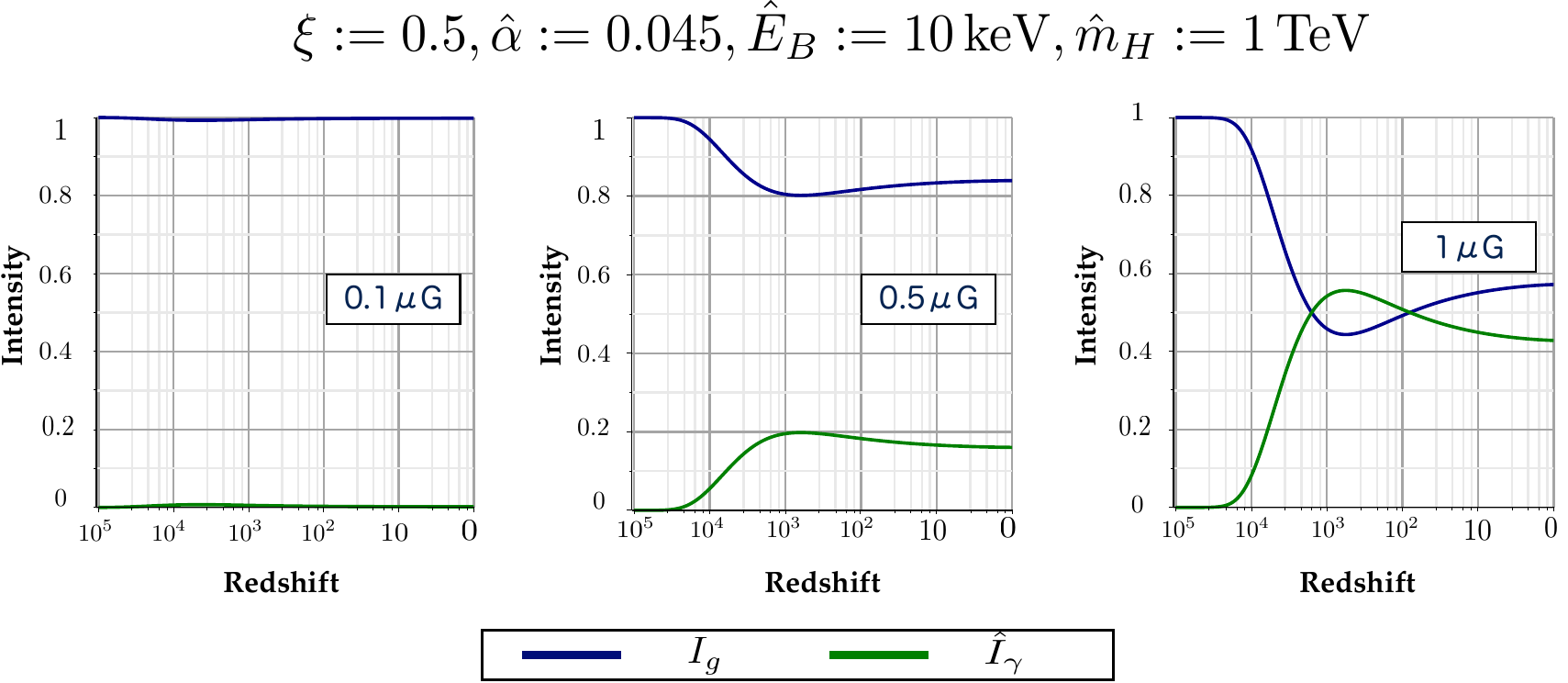}
	\caption{\label{FIG6}{\bf Smaller $\hat{\alpha}\ $($\omega_0 =1\,{\rm eV}$)}\\ The intensity of graviton and 
	dark photon is plotted as a function of redshift $z$, changing current dark magnetic field strength $\hat{B}_{0}$.}
\end{center}
\end{figure}
\begin{figure}[H]
\begin{center}
	\includegraphics[width=13cm]{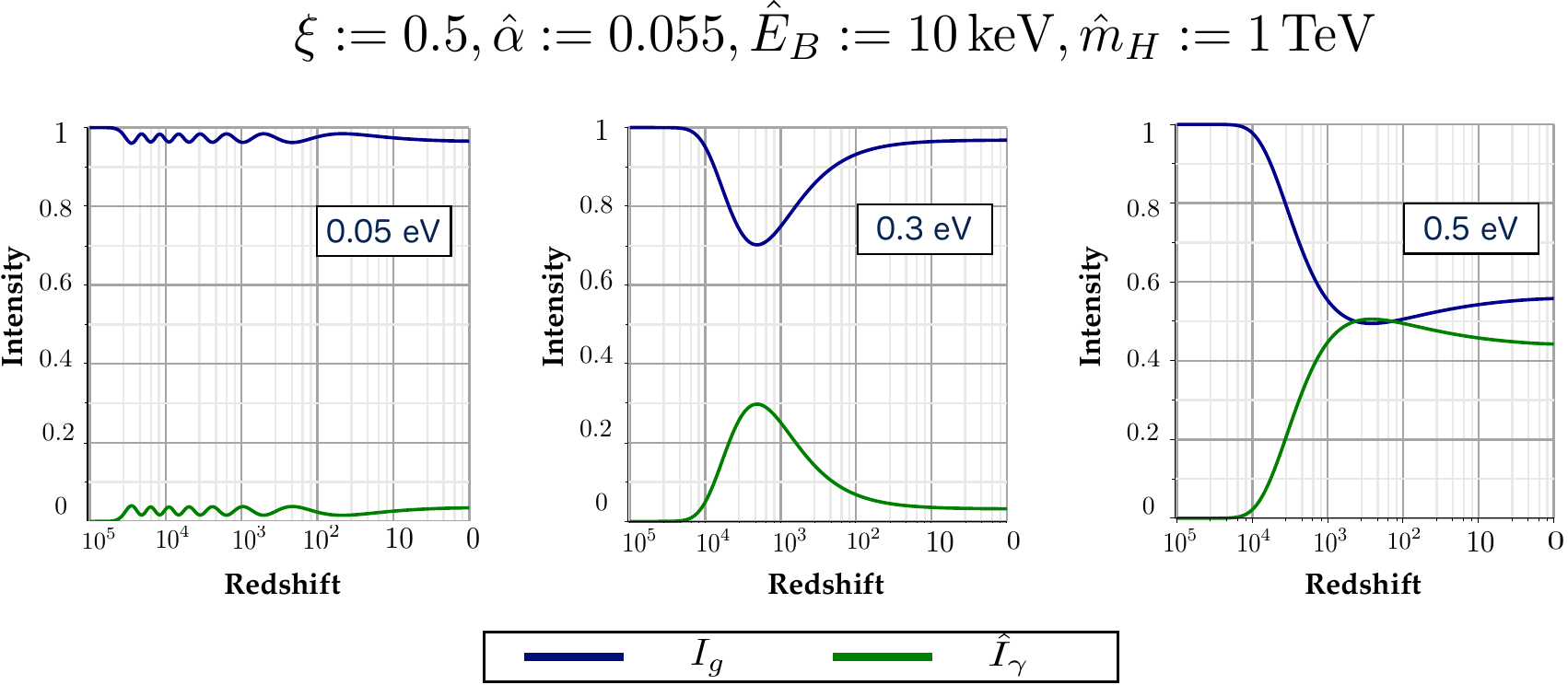}
	\caption{\label{FIG7}{\bf Bigger $\hat{\alpha}\ $($\hat{B}_{0} =1\,\mu{\rm G}$)}\\ The intensity of graviton and 
	dark photon is plotted as a function of redshift $z$, changing current gravitational waves energy $\omega_0$.}
\end{center}
\end{figure}
\begin{figure}[H]
\begin{center}
	\includegraphics[width=13cm]{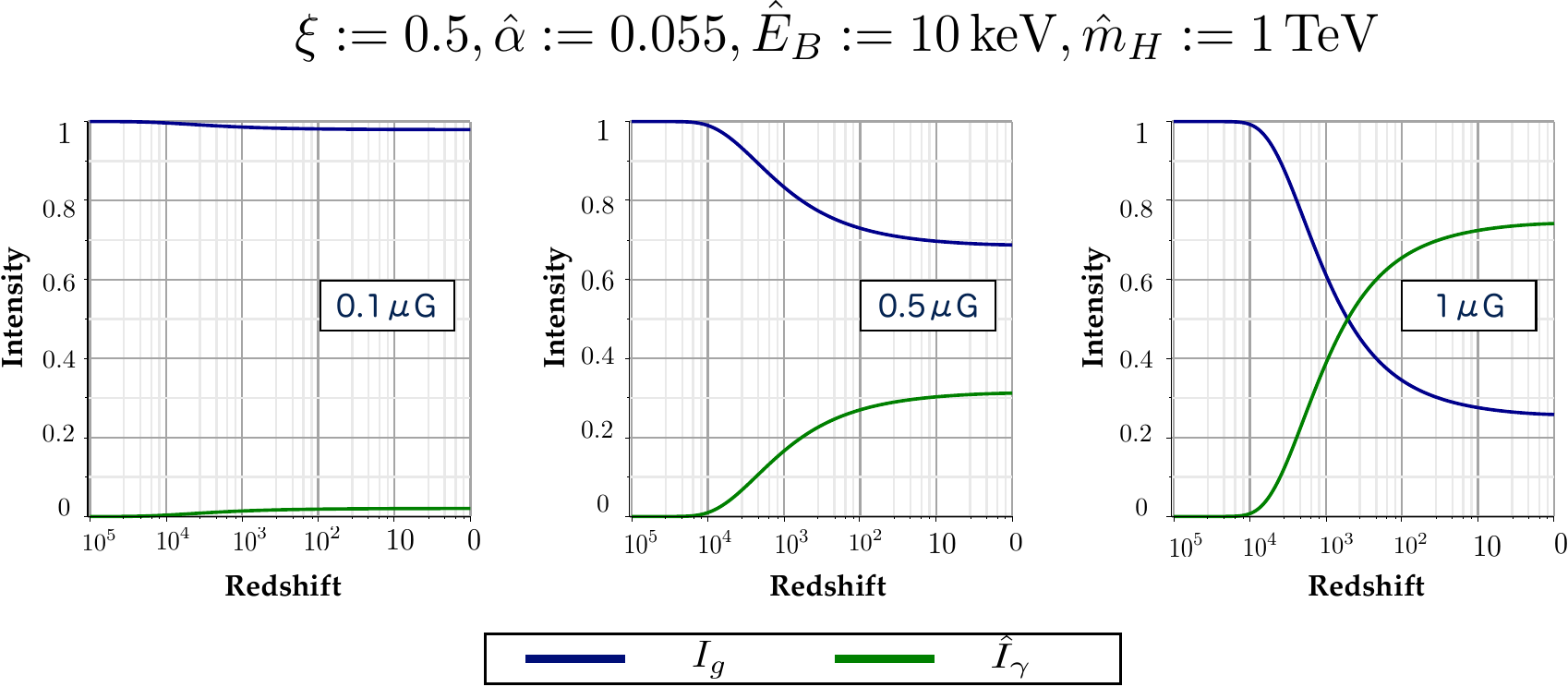}
	\caption{\label{FIG8}{\bf Bigger $\hat{\alpha}\ $($\omega_0 =1\,{\rm eV}$)}\\ The intensity of graviton and 
	dark photon is plotted as a function of redshift $z$, changing current dark magnetic field strength $\hat{B}_{0}$.}
\end{center}
\end{figure}

\subsection{Change in dark binding energy $\hat{E}_{\rm B}$}
\begin{figure}[H]
\begin{center}
	\includegraphics[width=13cm]{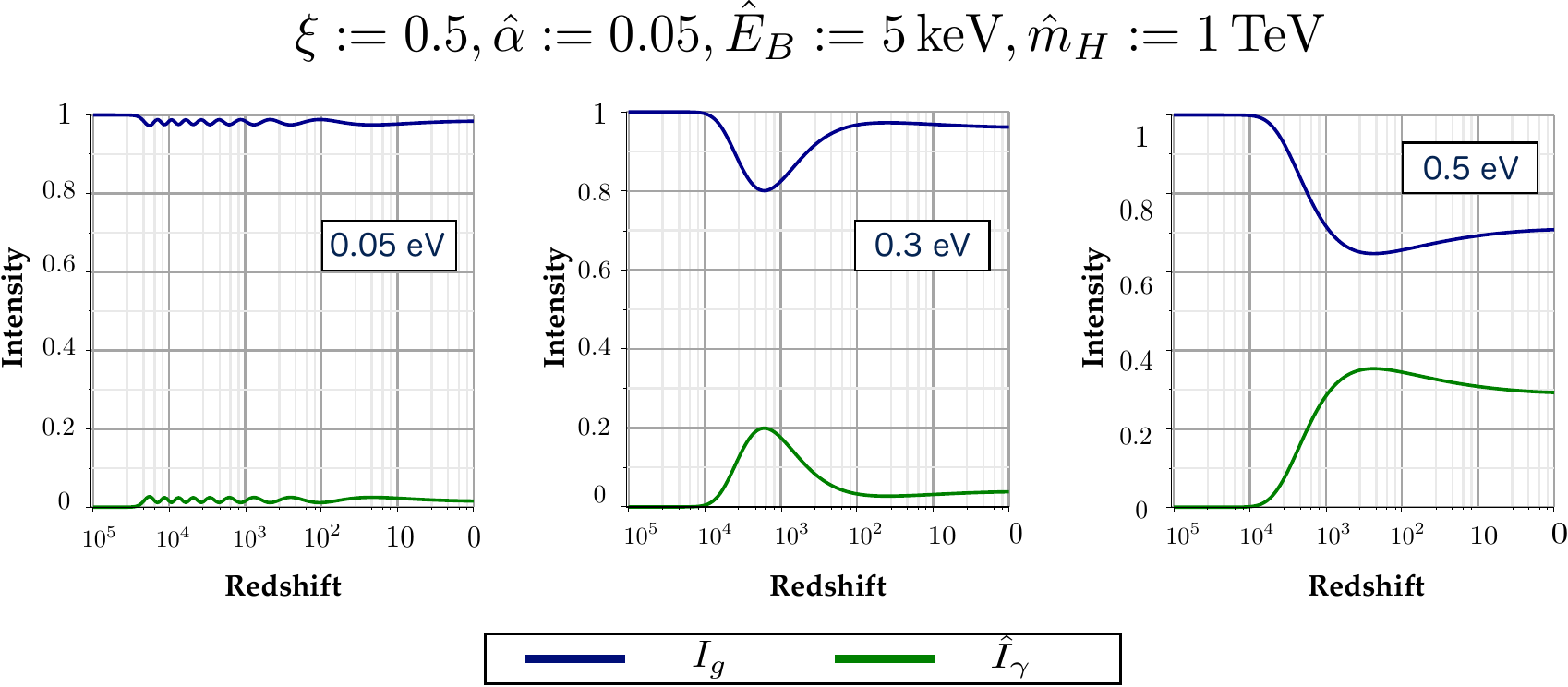}
	\caption{\label{FIG9}{\bf Smaller $\hat{E}_{\rm B}\ $($\hat{B}_{0} =1\,\mu{\rm G}$)}\\ The intensity of graviton and 
	dark photon is plotted as a function of redshift $z$, changing current gravitational waves energy $\omega_0$.}
\end{center}
\end{figure}

\begin{figure}[H]
\begin{center}
	\includegraphics[width=13cm]{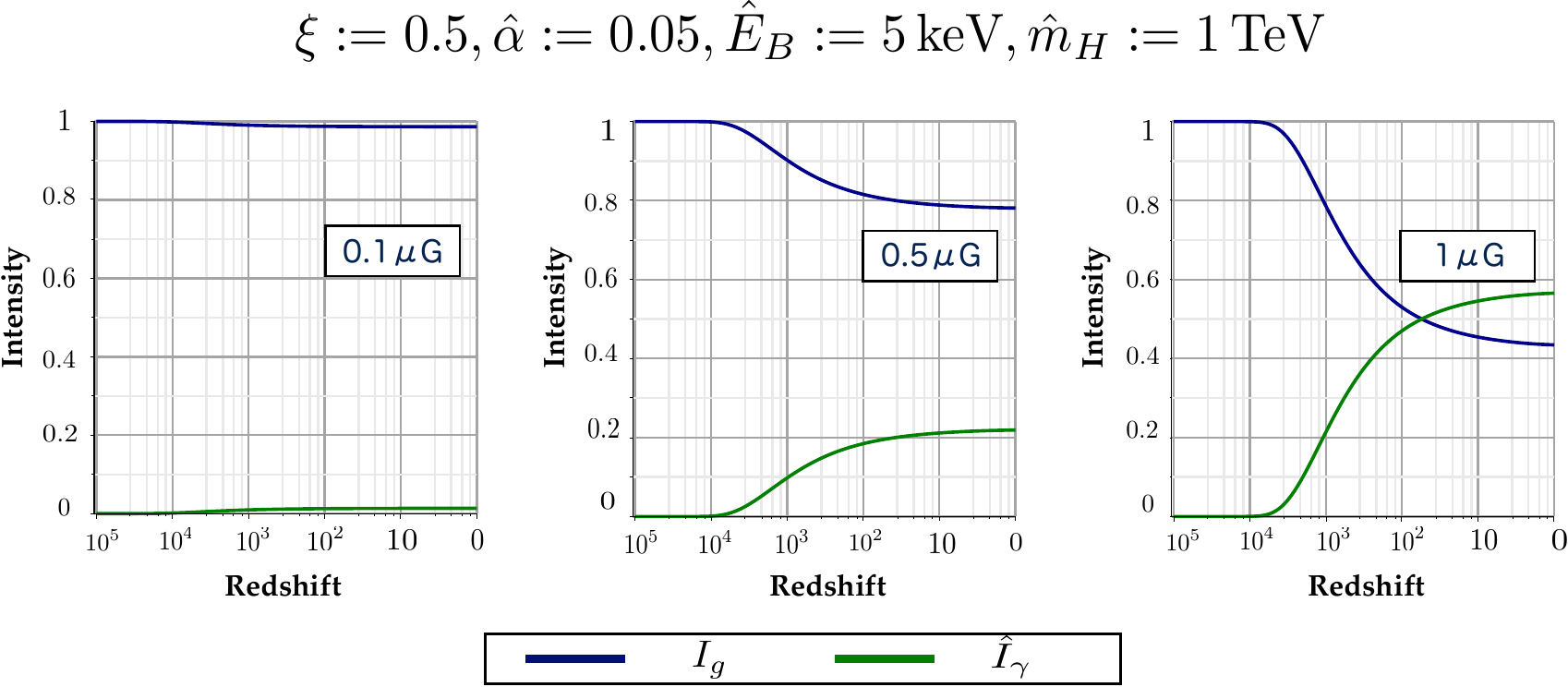}
	\caption{\label{FIG10}{\bf Smaller $\hat{E}_{\rm B}\ $($\omega_0 =1\,{\rm eV}$)}\\ The intensity of graviton and 
	dark photon is plotted as a function of redshift $z$, changing current dark magnetic field strength $\hat{B}_{0}$.}

\end{center}
\end{figure}

\begin{figure}[H]
\begin{center}
	\includegraphics[width=13cm]{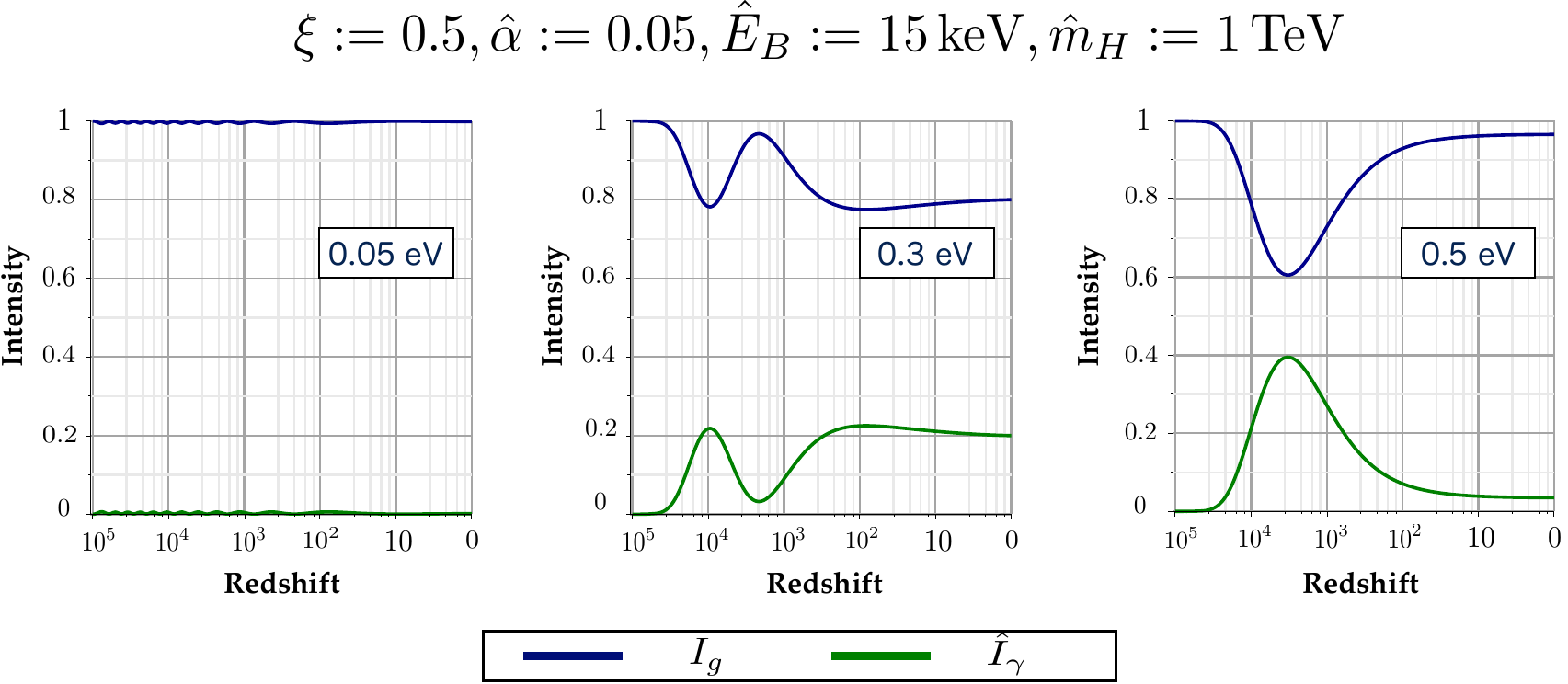}
	\caption{\label{FIG11}{\bf Bigger $\hat{E}_{\rm B}\ $($\hat{B}_{0} =1\,\mu{\rm G}$)}\\ The intensity of graviton and 
	dark photon is plotted as a function of redshift $z$, changing current gravitational waves energy $\omega_0$.}
\end{center}
\end{figure}

\begin{figure}[H]
\begin{center}
	\includegraphics[width=13cm]{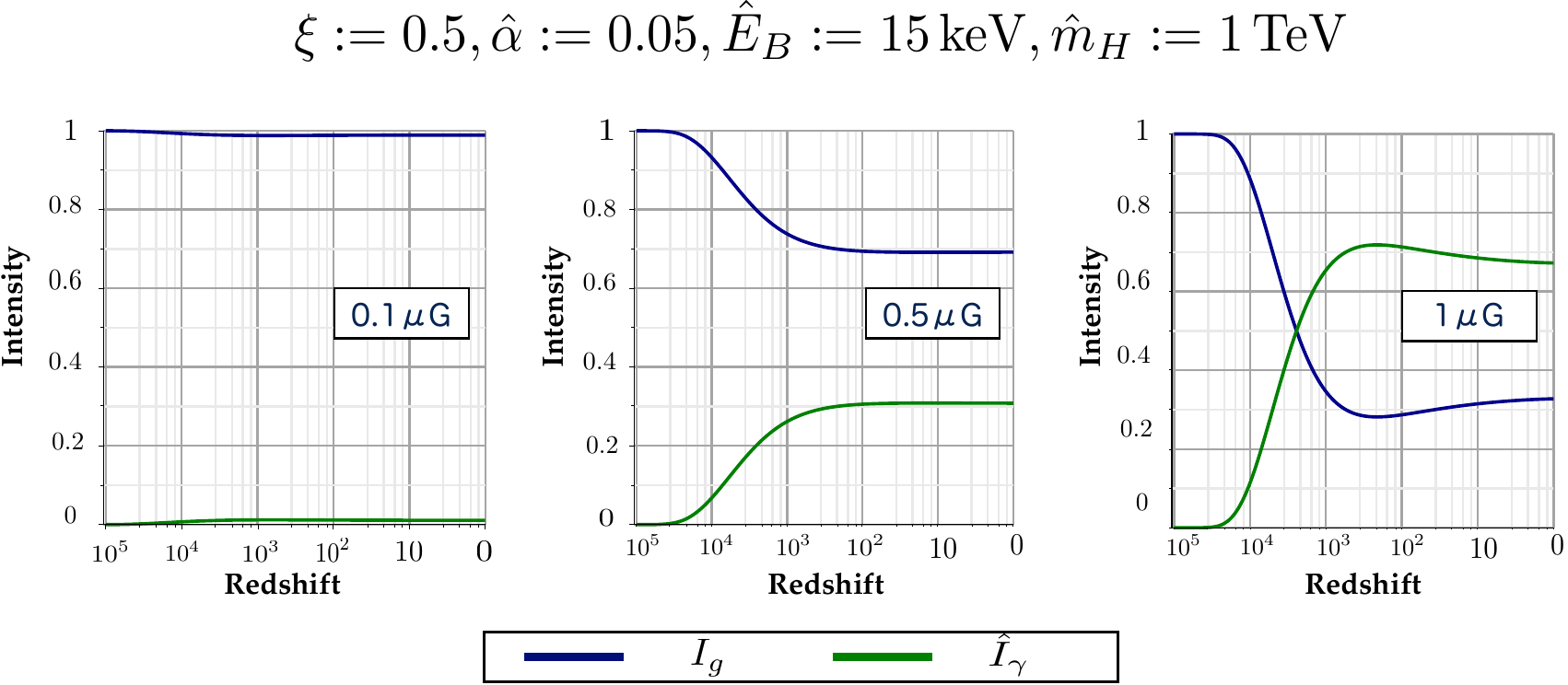}
	\caption{\label{FIG12}{\bf Bigger $\hat{E}_{\rm B}\ $($\omega_0 =1\,{\rm eV}$)}\\ The intensity of graviton and 
	dark photon is plotted as a function of redshift $z$, changing current dark magnetic field strength $\hat{B}_{0}$.}
\end{center}
\end{figure}

\subsection{Change in mass of dark hydrogen atom $\hat{m}_{\rm H}$}
\begin{figure}[H]
\begin{center}
	\includegraphics[width=13cm]{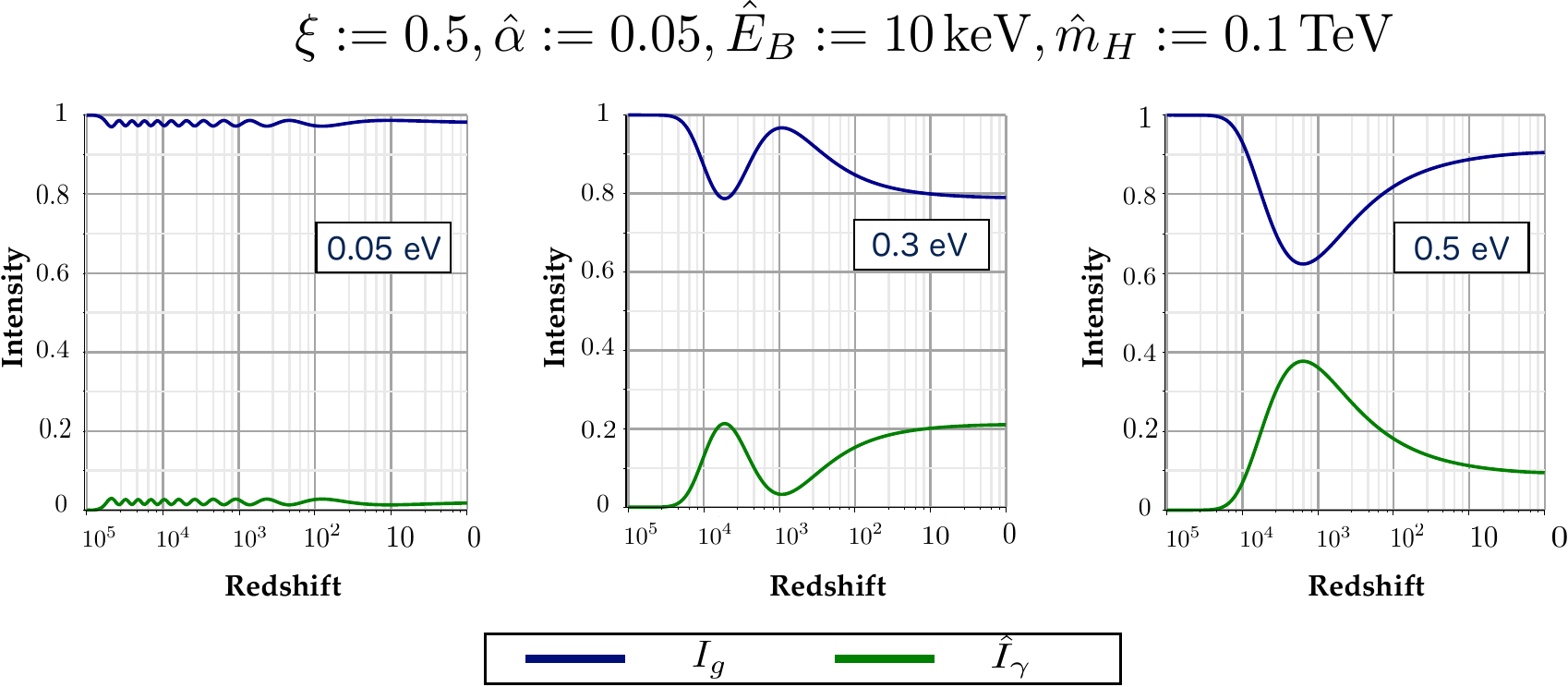}
	\caption{\label{FIG13}{\bf Smaller $\hat{m}_{\rm H}\ $($\hat{B}_{0} =1\,\mu{\rm G}$)}\\ The intensity of graviton and 
	dark photon is plotted as a function of redshift $z$, changing current gravitational waves energy $\omega_0$.}

\end{center}
\end{figure}

\begin{figure}[H]
\begin{center}
	\includegraphics[width=13cm]{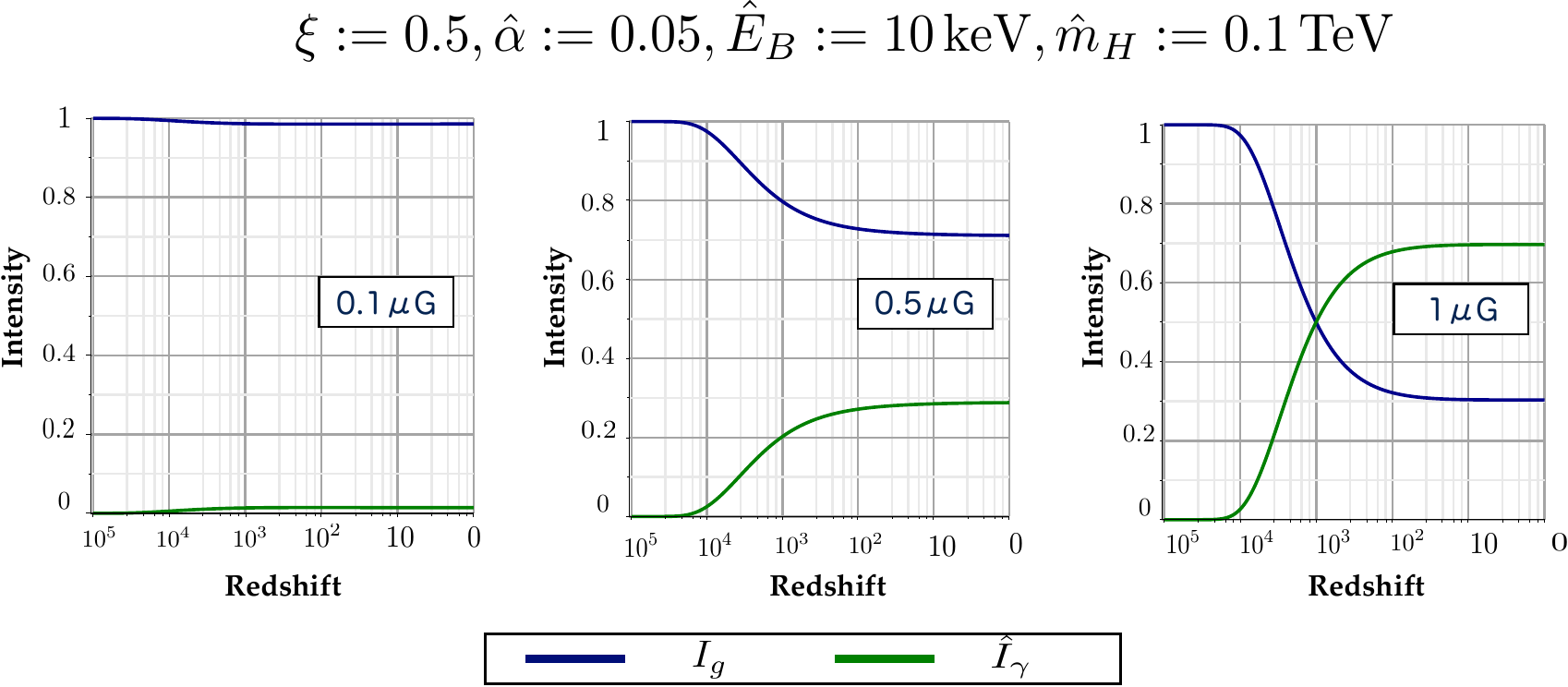}
	\caption{\label{FIG14}{\bf Smaller  $\hat{m}_{\rm H}\ $($\omega_0 =1\,{\rm eV}$)}\\ The intensity of graviton and 
	dark photon is plotted as a function of redshift $z$, changing current dark magnetic field strength $\hat{B}_{0}$.}
\end{center}
\end{figure}

\begin{figure}[H]
\begin{center}
	\includegraphics[width=13cm]{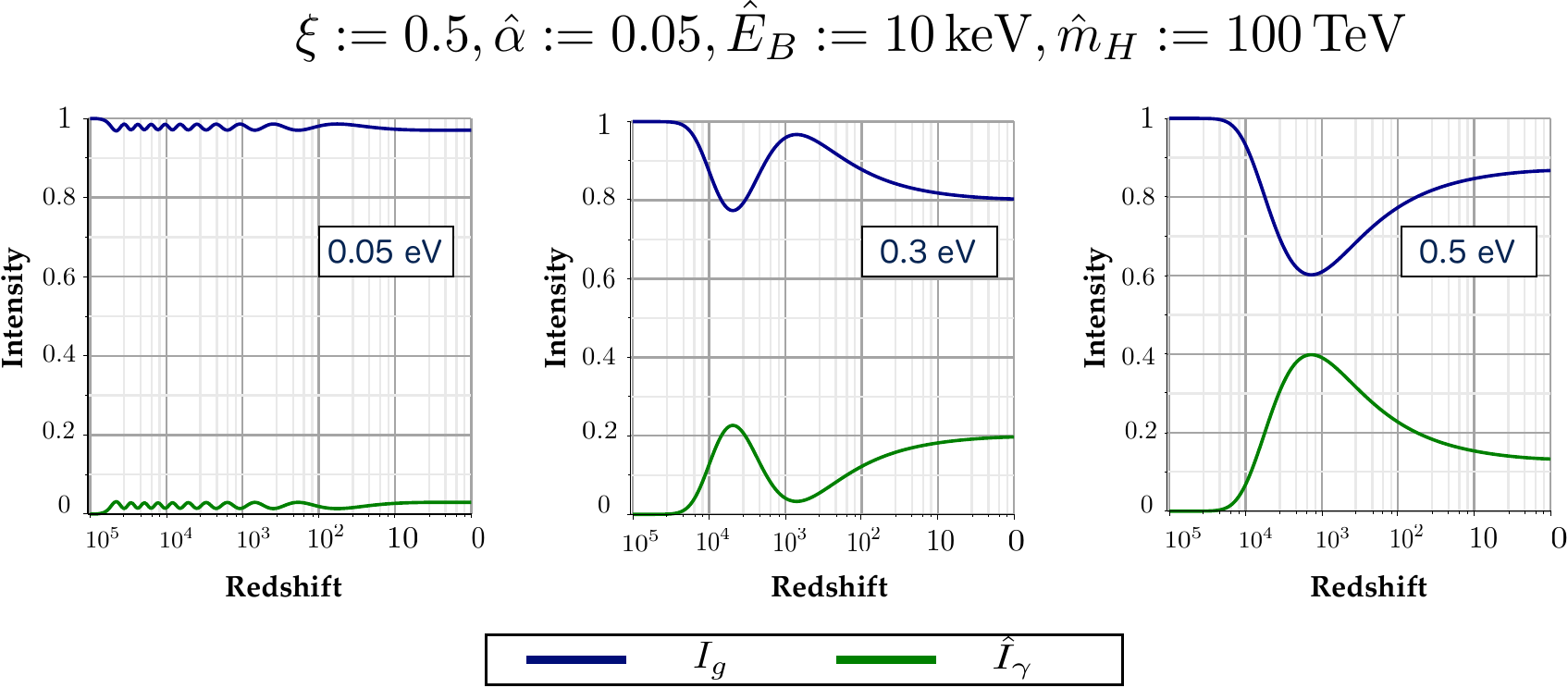}
	\caption{\label{FIG15}{\bf Bigger $\hat{m}_{\rm H}\ $($\hat{B}_{0} =1\,\mu{\rm G}$)}\\ The intensity of graviton and 
	dark photon is plotted as a function of redshift $z$, changing current gravitational waves energy $\omega_0$.}
\end{center}
\end{figure}

\begin{figure}[H]
\begin{center}
	\includegraphics[width=13cm]{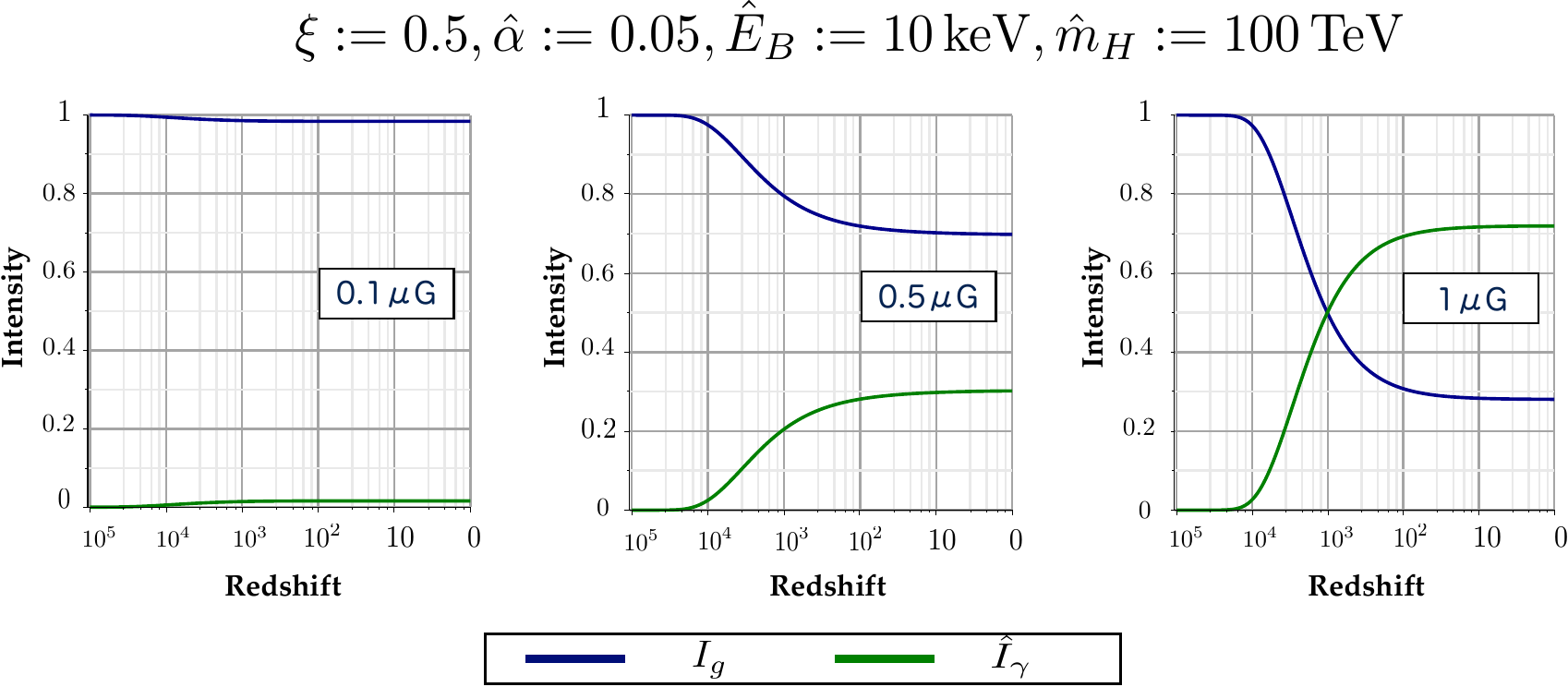}
	\caption{\label{FIG16}{\bf Bigger $\hat{m}_{\rm H}\ $($\omega_0 =1\,{\rm eV}$)}\\ The intensity of graviton and 
	dark photon is plotted as a function of redshift $z$, changing current dark magnetic field strength $\hat{B}_{0}$.}
\end{center}
\end{figure}

\subsection{Change in $\xi$}
\begin{figure}[H]
\begin{center}
	\includegraphics[width=13cm]{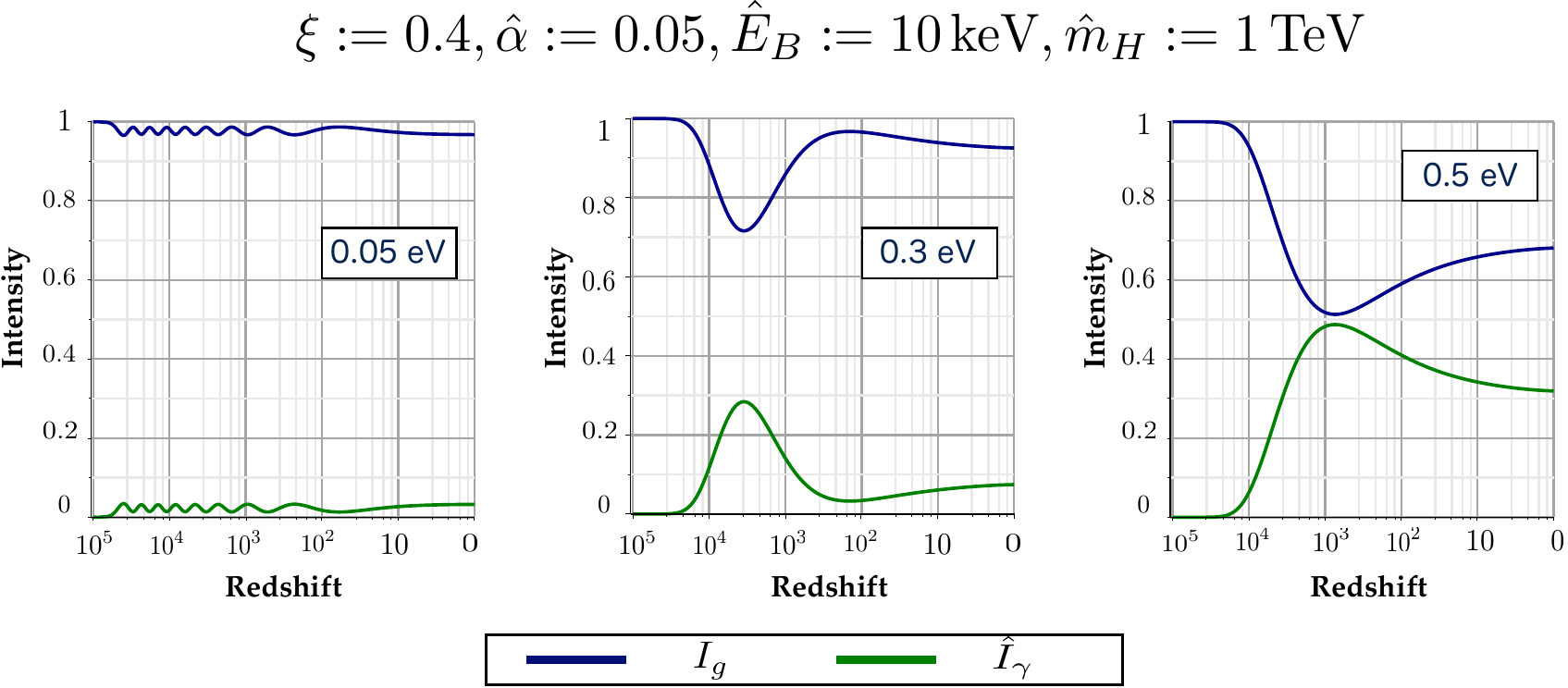}
	\caption{\label{FIG17}{\bf Smaller $\xi\ $($\hat{B}_{0} =1\,\mu{\rm G}$)}\\ The intensity of graviton and 
	dark photon is plotted as a function of redshift $z$, changing current gravitational waves energy $\omega_0$.}
\end{center}
\end{figure}

\begin{figure}[H]
\begin{center}
	\includegraphics[width=13cm]{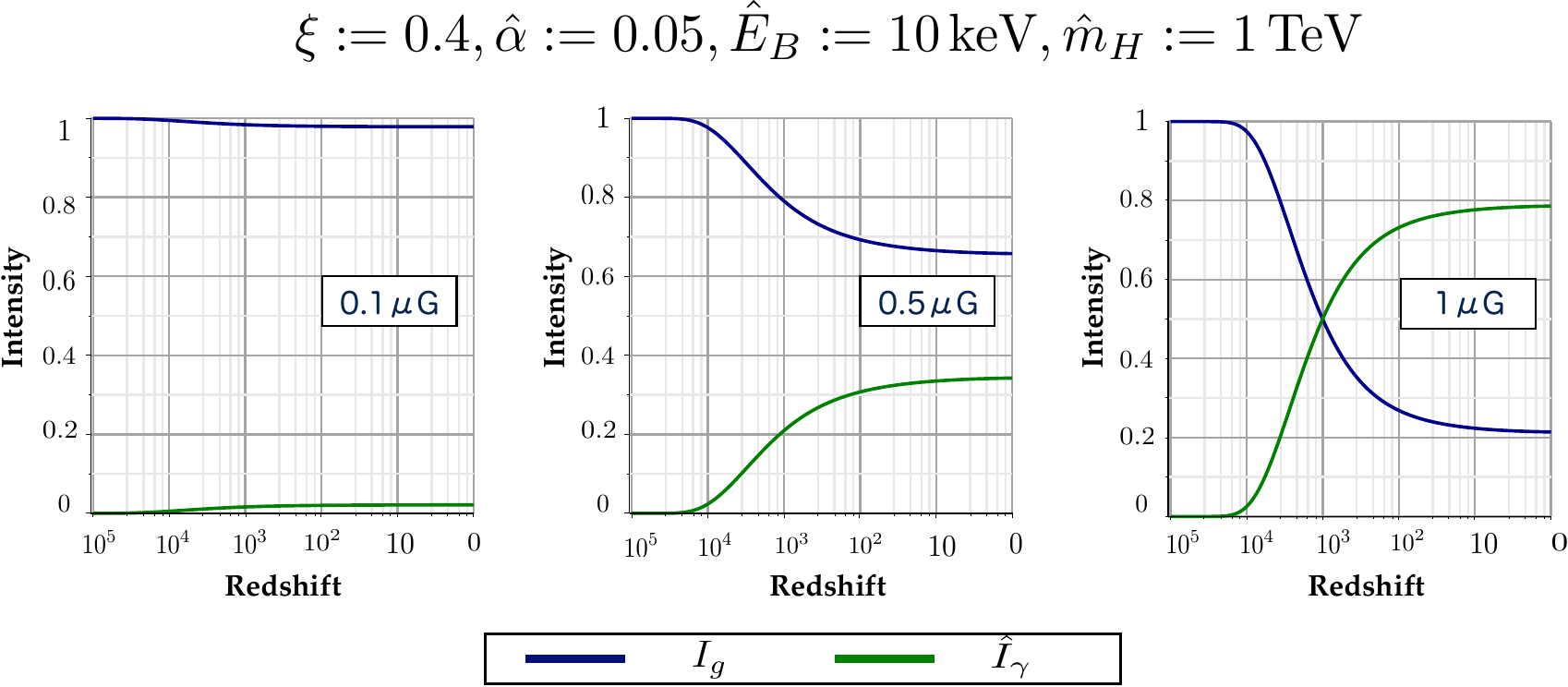}
	\caption{\label{FIG18}{\bf Smaller $\xi\ $($\omega_0 =1\,{\rm eV}$)}\\ The intensity of graviton and 
	dark photon is plotted as a function of redshift $z$, changing current dark magnetic field strength $\hat{B}_{0}$.}
\end{center}
\end{figure}

\begin{figure}[H]
\begin{center}
	\includegraphics[width=13cm]{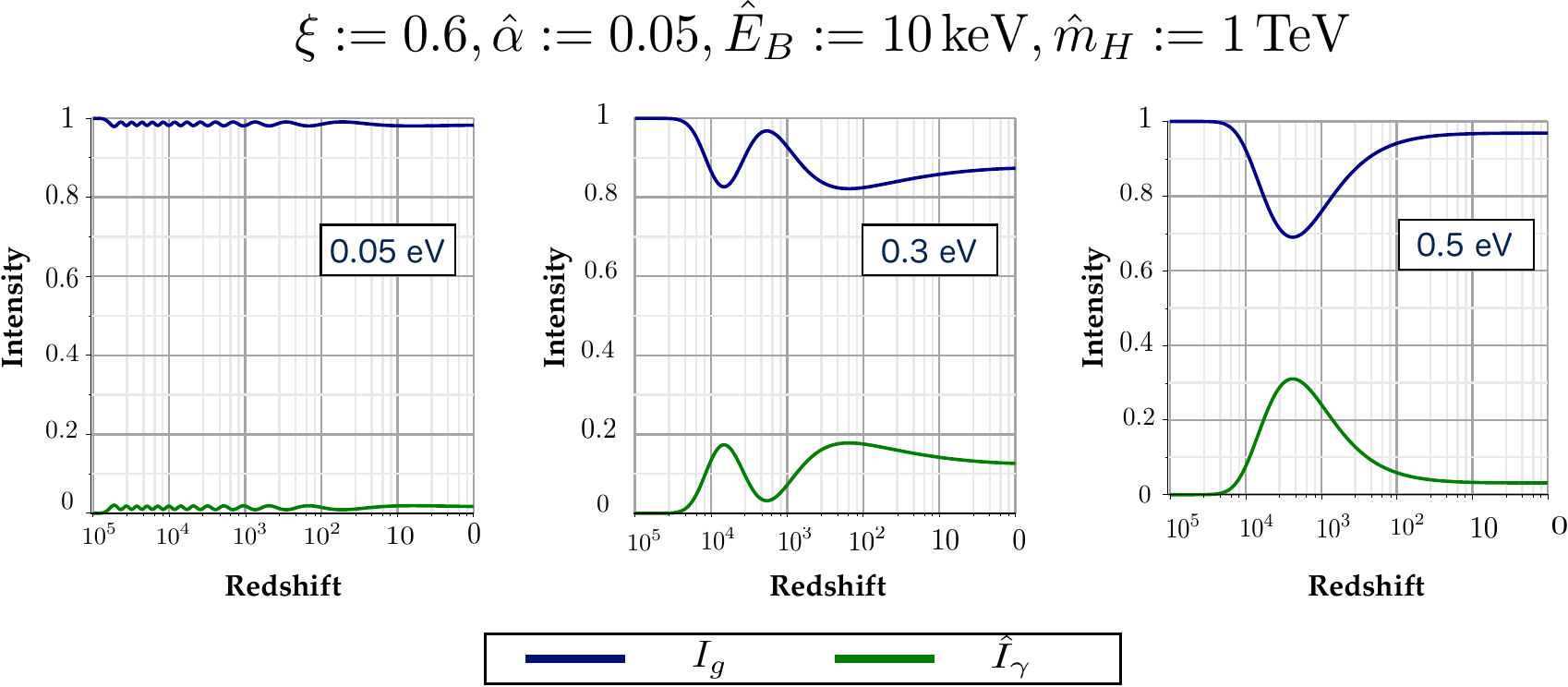}
	\caption{\label{FIG19}{\bf Bigger $\xi\ $($\omega_0 =1\,{\rm eV}$)}\\ The intensity of graviton and 
	dark photon is plotted as a function of redshift $z$, changing current gravitational waves energy $\omega_0$.}
\end{center}
\end{figure}

\begin{figure}[H]
\begin{center}
	\includegraphics[width=13cm]{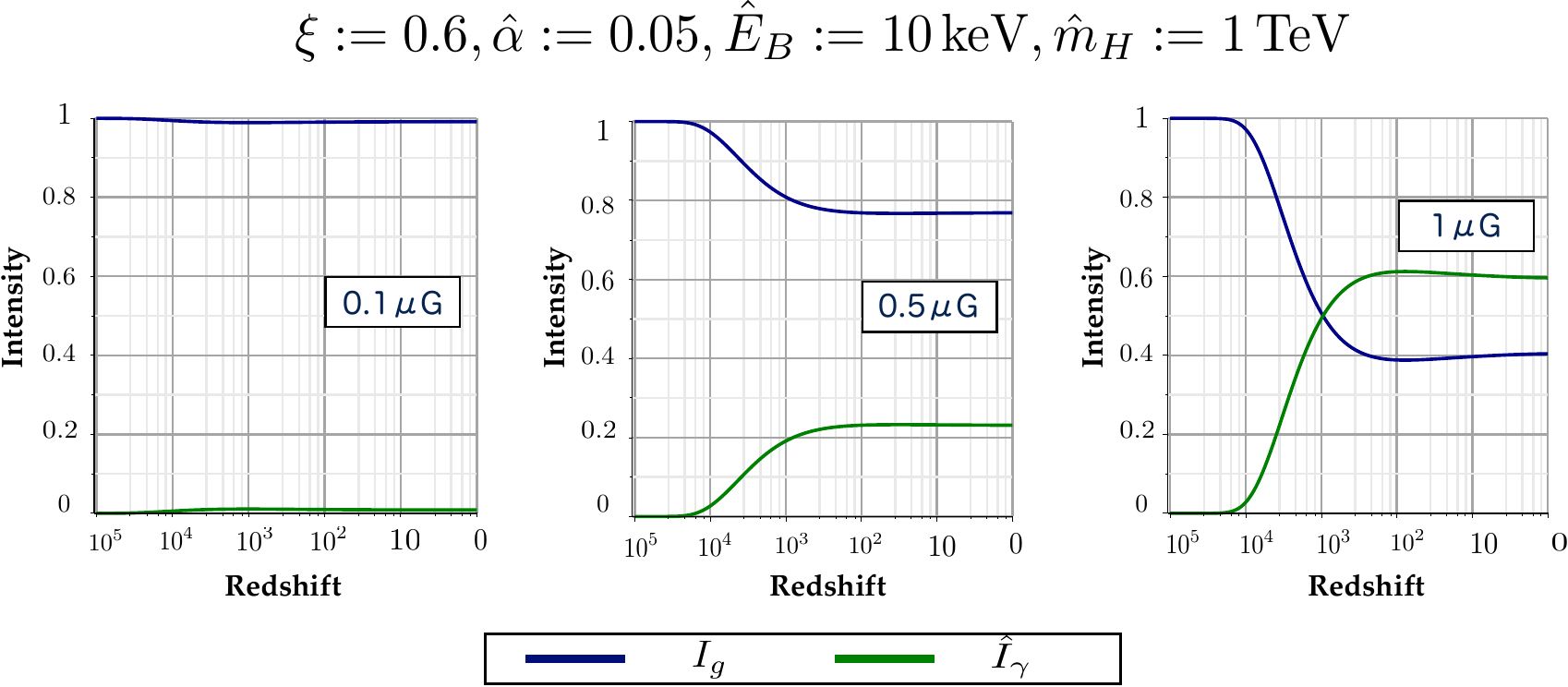}
	\caption{\label{FIG20}{\bf Bigger $\xi\ $($\hat{B}_{0} =1\,\mu{\rm G}$)}\\ The intensity of graviton and 
	dark photon is plotted as a function of redshift $z$, changing current dark magnetic field strength $\hat{B}_{0}$.}

\end{center}
\end{figure}

\newpage
\nocite{*}

\bibliography{DR12127_revised_proof1.bib}
\end{document}